\documentclass{article}
\usepackage{bm,latexsym,amsmath,amssymb,amsfonts,fancyhdr,color,graphicx,multirow,slashed,cite}
\usepackage[a4paper,bottom=3cm,top=2.5cm,head=0mm,width=17cm,dvipdfm]{geometry}
\usepackage[usenames,dvipsnames,svgnames,table]{xcolor}
\usepackage[colorlinks=true,
            linkcolor=blue,
            urlcolor=blue,
            citecolor=green,          
						bookmarks=true,
						bookmarksnumbered=true,
						breaklinks=true,
						pdfpagemode=Fullscreen,
						pdfstartview=FitBH]{hyperref}
\usepackage[dotinlabels]{titletoc}
\usepackage{titlesec}
\usepackage{authblk,ulem}

\numberwithin{equation}{section}
\allowdisplaybreaks[4]

\titlelabel{\thetitle.\quad \hspace{-0.8em}}
\titlecontents{section}
              [1.5em]
              {\vspace{4mm} \large \bf}
              {\contentslabel{1em}}
              {\hspace*{-1em}}
              {\titlerule*[.5pc]{.}\contentspage}
\titlecontents{subsection}
              [3.5em]
              {\vspace{2mm}}
              {\contentslabel{1.8em}}
              {\hspace*{.3em}}
              {\titlerule*[.5pc]{.}\contentspage}
\titlecontents{subsubsection}
              [5.5em]
              {\vspace{2mm}}
              {\contentslabel{2.5em}}
              {\hspace*{.3em}}
              {\titlerule*[.5pc]{.}\contentspage}

\newcommand{\titledef}{Associated Production of Neutrino and Dark Fermion at Future Lepton Colliders} 

\hypersetup{ pdfauthor = {Shao-Feng Ge},
	     pdftitle = {\titledef}, 
	     pdfsubject = {}, 
             pdfkeywords = {}, 
	     pdfcreator = {LaTeX with hyperref package},
	     pdfproducer = {dvips + ps2pdf} }

\definecolor{gesfblack}{rgb}{0,0,0}

\definecolor{gesfblue}{rgb}{0.08,0.42,0.76}

\definecolor{gesfgreen}{rgb}{0,1,0}

\definecolor{gesfgrey}{rgb}{0.5,0.5,0.5}

\definecolor{gesflanse}{rgb}{0.00,0.50,0.50}

\definecolor{gesfpurple}{rgb}{0.47,0.19,0.42}

\definecolor{gesfred}{rgb}{1,0,0}

\definecolor{gesfwhite}{rgb}{1,1,1}

\definecolor{gesfyellow}{rgb}{0.7,0.4,0.3}

\newcommand{\gsec}[1]{{\hypersetup{linkcolor=red}Sec.\,\ref{#1}\hypersetup{linkcolor=blue}}}

\newcommand{\geqn}[1]{\hypersetup{linkcolor=blue}Eq.\,(\ref{#1})\hypersetup{linkcolor=blue}}
\newcommand{\gfig}[1]{{\hypersetup{linkcolor=violet}Fig.\,\ref{#1}\hypersetup{linkcolor=blue}}}
\newcommand{\gtab}[1]{{\hypersetup{linkcolor=gesflanse}Table~\ref{#1}\hypersetup{linkcolor=blue}}}

\definecolor{Orange}{cmyk}{0,0.61,0.87,0}
\definecolor{JungleGreen}{cmyk}{0.99,0,0.52,0}
\definecolor{OliveGreen}{cmyk}{0.64,0,0.95,0.40}
\definecolor{Brown}{cmyk}{0,0.81,1,0.60}
\definecolor{RoyalBlue}{cmyk}{0.71,0.53,0,0.12}
\definecolor{Gray}{cmyk}{0,0,0,0.40}
\definecolor{LightPink}{cmyk}{0.0,0.25,0,0}
\definecolor{LLightPink}{cmyk}{0.0,0.10,0,0}
\definecolor{LightBlue}{cmyk}{0.25,0,0,0}
\definecolor{LightGray}{cmyk}{0,0,0,0.2}

\graphicspath{{figs/}}

\usepackage{subfigure}
\newcommand{\bee}{\begin{equation}}
\newcommand{\ene}{\end{equation}}
\newcommand{\bea}{\begin{eqnarray}}
\newcommand{\ena}{\end{eqnarray}}

\def\ab{\, {\rm ab}}
\def\tev{\,{\rm TeV}}
\def\gev{\,{\rm GeV}}

\def\call{\mathcal{L}}

\def\calo{\mathcal{O}}

\def\ab{\, {\rm ab}}
\def\met{{E\!\!\!\slash}_{T} }

\def\ie{{\it i.e.}}

\begin{document}
\fontsize{12pt}{14pt}\selectfont

\title{
       \bf \titledef} 
\author[1,2]{{\large Shao-Feng Ge} \footnote{\href{mailto:gesf@sjtu.edu.cn}{gesf@sjtu.edu.cn}}}
\affil[1]{Tsung-Dao Lee Institute \& School of Physics and Astronomy, Shanghai Jiao Tong University, Shanghai 200240, China}
\affil[2]{Key Laboratory for Particle Astrophysics and Cosmology (MOE) \& Shanghai Key Laboratory for Particle Physics and Cosmology, Shanghai Jiao Tong University, Shanghai 200240, China}
\author[3,4]{{\large Kai Ma} \footnote{\href{mailto:makai@ucas.ac.cn}{makai@ucas.ac.cn}}}
\affil[3]{Department of Physics, Shaanxi University of Technology, Hanzhong 723000, Shaanxi, China}
\affil[4]{
Ministry of Education Key Laboratory for Nonequilibrium Synthesis and Modulation of Condensed Matter\\
Shaanxi Province Key Laboratory of Quantum Information and Quantum Optoelectronic Devices\\ 
School of Physics, Xi'an Jiaotong University, Xi'an 710049, China
}
\author[5,6,7]{{\large Xiao-Dong Ma}
\footnote{\href{mailto:maxid@scnu.edu.cn}{maxid@scnu.edu.cn}}}
\affil[5]{Key Laboratory of Atomic and Subatomic Structure and Quantum Control (MOE), Institute of Quantum Matter, South China Normal University, Guangzhou 510006, China}
\affil[6]{Guangdong Provincial Key Laboratory of Nuclear Science, Institute of Quantum Matter, South China Normal University,
Guangzhou 510006, China}
\affil[7]{Guangdong-Hong Kong Joint Laboratory of Quantum Matter,
Southern Nuclear Science Computing Center, South China Normal University, Guangzhou 510006, China}
\author[1,2]{{\large Jie Sheng} \footnote{\href{mailto:shengjie04@sjtu.edu.cn}{shengjie04@sjtu.edu.cn}}}
\date{}

\maketitle

\vspace{-2mm}
\begin{abstract}
\fontsize{10pt}{12pt}\selectfont
Fermionic dark matter can be pairly produced and hence
searched with missing energy at colliders. We extend
such probe to the associated production of a neutrino and
a dark sector fermion at the future $e^+ e^-$ colliders
such as CEPC, FCC-ee, ILC, and CLIC. 
Two typical processes, the mono-photon and electron-positron pair
productions associated with missing energy, can serve the
purpose. While the mono-photon search prevails at CEPC, FCC-ee,
and ILC, the $e^+ e^- \met$ channel has more significant contributions
at CLIC with much higher collision energy $\sqrt s$. The beam polarizations
can help further suppressing the SM backgrounds to enhance the signal
significance while differential cross sections can
distinguish the Lorentz structure of various effective operators.
The combined sensitivity can reach well above $1\tev$ at
CEPC/FCC-ee and ILC while it further touches 30\,TeV at CLIC.
Comparing with the updated results from the direct
detection experiments (XENON1T, PandaX-II, PandaX-4T, LZ,
and XENONnT), astrophysical $X/\gamma$-ray observations,
and cosmological
constraints for the sub-MeV absorption dark matter,
the collider searches are actually more sensitive
and hence can provide a complementary approach to addressing the
dark fermions.
\end{abstract}

\section{Introduction}
\label{sec:intro}

The existence of dark matter (DM) has been established by various
cosmological and astrophysical observations
\cite{Bertone:2004pz,Young:2016ala,Arbey:2021gdg}.
However, its identity and physical properties are still unknown. 
We need more observational clues for the DM model
building \cite{Roszkowski:2017nbc}. In addition to the
cosmological searches \cite{Valluri:2022nrh},
there are mainly three types of DM searches: direct detection \cite{Liu:2017drf,Schumann:2019eaa,Billard:2021uyg}, 
indirect detection \cite{Gaskins:2016cha,Leane:2020liq,Slatyer:2021qgc},
and collider search \cite{Boveia:2018yeb,Gori:2022vri,Lagouri:2022ier,Penning:2017tmb},
to constrain the couplings between the DM and SM
particles \cite{Roszkowski:2017nbc,Bottaro:2021snn,Bertone:2018krk}.
Most of them are given on the DM coupling with nucleons and
hence effectively quarks.
Note that colliders probe not just
the true DM particle that survives until today
but also any dark sector particles that
can be directly generated as long as it is kinematically allowed.
However, no clear evidence has been
observed yet and hence puts strong constraints.

So it seems promising for DM to have leptophilic interactions.
For instance, if the DM only directly couples with leptons,
its coupling with quarks is then suppressed.  
Such leptophilic scenarios are characterized by a conspicuous gamma-ray 
in the galactic halo because of the 
radiatively induced annihilation rate into leptons and 
photons \cite{Baltz:2002we,John:2021ugy,Bi:2009uj, Ibarra:2009bm}. One particularly interesting possibility
is the sterile neutrino DM that can produce astrophysical $X$-ray
\cite{Davis:2015vla,Drewes:2016upu,Kouvaris:2016afs,Abazajian:2017tcc,Boyarsky:2018tvu,Kopp:2021jlk}.
The leptophilic DM models have been discussed extensively in literature~\cite{Fox:2008kb,Cao:2009yy,Davoudiasl:2009dg,Kopp:2009et,Cohen:2009fz,Chun:2009zx,Haba:2010ag,Ko:2010at,Schmidt:2012yg,Das:2013jca,Chang:2014tea,Bell:2014tta,Cao:2014cda,Chao:2017emq,Ghorbani:2017cey,Han:2017ars,Madge:2018gfl,Horigome:2021qof,Ghosh:2020fdc}
and have the potential of explaining various experimental anomalies
\cite{Kundu:2021cmo,Barman:2021hhg,Liang:2021kgw}.

The collider search provides a tunable environment to distinguish
the leptophilic and hadrophilic natures of DM. While hadron colliders
mainly probe quark couplings
\cite{delAguila:2014soa,Buckley:2015cia,Farzan:2010mr,Su:2009fz,Kahlhoefer:2017dnp}, 
lepton colliders are more sensitive to the leptonic ones
\cite{Bartels:2012ex,Kundu:2021cmo,Barman:2021hhg,Liang:2021kgw,Freitas:2014jla,Dreiner:2012xm,Habermehl:2020njb,Kalinowski:2021tyr,Bharadwaj:2020aal}.
Typically, the DM search at lepton collider utilizes the missing
energy or momentum as characteristic signature, such as the mono-photon
\cite{Fox:2011fx,Chae:2012bq,DELPHI:2003dlq,DELPHI:2008uka,Birkedal:2004xn,Liu:2019ogn}
and mono-$Z$ \cite{Wan:2014rhl,Yu:2014ula,Dutta:2017ljq,Grzadkowski:2020frj} 
signals that have been studied extensively. In both cases, the associated
production of a single bosonic DM or a pair of fermionic DM particles
carries away missing energy
\cite{Albert:2016osu,Arun:2017uaw,Lin:2019uvt,Chen:2008dh}.
This does not exhaust all the possibilities.

In addition to DM that survives in the Universe, neutrino and other dark fermions
can also carry away some missing energy. It is
natural to explore the possible associated production
of dark fermion $\chi$ and neutrino $\nu$. This coincides with the fermionic
DM absorption on electron target,
$\chi e \to \nu e$ in the sub-MeV mass range. Putting both electron legs to the initial
state, a single dark fermion can also appear in the
final state from the $e^+ e^-$ annihilation at lepton colliders.
For instance, the process $e^+e^- \to \chi \nu$ can happen
as long as the collision energy $\sqrt{s} > m_\chi$.
One may also expect such collider searches
to provide complementary constraints on the same operator as
explored in the direct detection at least in the
sub-MeV mass range. For the remaining mass range
above MeV, the future lepton colliders serves as unique
probe of other dark fermions.

The rest of this paper is organized as follows.
In \gsec{sec:absorptionDM}, we summarize the profound
motivation and essential features of the fermionic absorption
DM on electron target. The next
\gsec{sec:monoa} studies the mono-photon production 
at the future $e^+e^-$ colliders with both polarized and
unpolarized beams
to obtain the expected exclusion limits at collision
energies of 240\gev, 500\gev, and 3\tev~for CEPC/FCC-ee,
ILC, and CLIC, respectively. We notice that
the $e^+e^-\met$ signal provides better sensitivity
at higher energies as elaborated in \gsec{sec:epemx}. 
In \gsec{sec:sec5}, we provide the updated constraints
from the DM overproduction, the astrophysical
$X/\gamma$-ray observations, and the direct detection
experiments for the sub-MeV fermionic absorption DM.
The fully combined results and our conclusions can be found
in \gsec{sec:conclusion}.

\section{Fermionic DM Absorption on Electron Target}
\label{sec:absorptionDM}

The direct detection of DM employs nuclear recoil to manifest
the existence of DM. Typically, a DM particle transfers
part of its kinetic energy to the target nuclei via elastic
scattering. Since the DM velocity has fixed distribution
by the gravitational potential in our Milky Way Galaxy, the
momentum transfer scales with the DM mass squared.
Using heavy nuclei as target, the direct detection experiment
is intrinsically sensitive to the weak scale DM particles.
In addition,
the energy threshold of direct detection experiments also limits the
sensitivity for light DM candidates.

These difficulties can be overcome if DM converts its
mass into recoil energies through the absorption process,
such as $\chi e \rightarrow \nu e$ for fermionic DM $\chi$.
Since the final-state neutrino is almost massless, the
DM mass is fully released. The aforementioned two problems,
recoil energy scales with DM mass squared and the sensitivity
is limited by the experimental energy threshold, can be
evaded simultaneously. Such direct detection process
can appear in a variant way at lepton colliders by moving
the two electron legs to the initial state,
$e^+ e^- \rightarrow \chi \bar \nu$ or $\bar \chi \nu$,
as we elaborate in this paper. More generally,
the collider search covers not just the genuine DM particle
that survives until today but also other dark fermions.
In addition to being complementary, the collider search
has its own advantage of exploring the whole dark sector.

Since the momentum transfer for the direct detection process
is typically small, it is enough to use effective
operators when exploring sensitivities there \cite{Ge:2022ius}.
In effective field theory (EFT), operators are usually 
constructed according to the SM
$SU(3)_c \times SU(2)_L \times U(1)_Y$ gauge symmetries.
However, the weak symmetry $SU(2)_L$ should have already been broken
at low energy and only the electromagnetic $U(1)_{\rm em}$
symmetry remains to constrain the form of effective operators.
The $\chi e \rightarrow \nu e$ and
$e^+ e^- \rightarrow \chi \bar \nu$ or $\bar \chi \nu$
process involves four fermions in the external states and
hence can be described by dimension-6 (dim-6) four-fermion effective
operators,
\begin{equation}
\begin{aligned} 
\mathcal{O}_{S} & \equiv (\bar{e} e)\left(\bar{\nu}_{L} \chi_{R}\right) \,,
\\ 
\mathcal{O}_{P} & \equiv \left(\bar{e} i \gamma_{5} e\right)\left(\bar{\nu}_{L} \chi_{R}\right) \,,
\\ 
\mathcal{O}_{V} & \equiv \left(\bar{e} \gamma_{\mu} e\right)\left(\bar{\nu}_{L} \gamma^{\mu} \chi_{L}\right) \,,
\\ 
\mathcal{O}_{A} & \equiv \left(\bar{e} \gamma_{\mu} \gamma_{5} e\right)\left(\bar{\nu}_{L} \gamma^{\mu} \chi_{L}\right) \,,
\\ 
\mathcal{O}_{T} & \equiv \left(\bar{e} \sigma_{\mu \nu} e\right)\left(\bar{\nu}_{L} \sigma^{\mu \nu} \chi_{R}\right) \,,
\end{aligned}
\label{eq:operators}
\end{equation}
as well as their hermitian conjugates \cite{Ge:2022ius}. 
The above parameterization is complete by including all the
five independent Lorentz structures for the electron bilinear.
Any other dim-6 operator can be expressed as a linear combination
of these 5 operators via Dirac $\gamma$ matrix identities and Fierz 
transformations \cite{Nieves:2003in,Nishi:2004st,Liao:2012uj}.
The neutrino is assumed to be left-handed while the dark fermion
$\chi$ is a Dirac fermion which has both left- and right-handed
components. In the effective Lagrangian,
\begin{equation}
  \mathcal L_{\mathrm{eff}}
=
  \sum_i \frac 1 {\Lambda^2_i} \mathcal O_i
+ \text {h.c.},
\end{equation}
each operator has a corresponding cut-off scale $\Lambda_i$
that is related to possible fundamental physics.
Although these absorption operators are originally
established for the fermionic absorption DM , they also apply
for any other dark fermion like sterile neutrino. We keep ``{\it absorption}''
in the name just as a reminder.

For collider searches, the EFT can still work as long as the cut-off
scale $\Lambda_i$ is higher than the collision energy.
As explored at PandaX-4T, the existing data can already constrain
the cut-off scale to be above 1\,TeV \cite{PandaX:2022ood}.
It is safe to follow the EFT approach for future
lepton colliders, such as CEPC/FCC-ee
at $\sqrt s = 240$\,GeV and ILC at
500\,GeV, respectively. Even for CLIC with $\sqrt s = 3$\,TeV,
the EFT approach can still work with even higher reach of the probed
scale, typically $\Lambda_i \sim \mathcal O(10)$\,TeV.

\section{Mono-Photon Production}
\label{sec:monoa}

\begin{figure}[ht]
\centering
\subfigure[]{
\label{fig:nca:Feyn:isr}
\includegraphics[height=0.18\textheight]{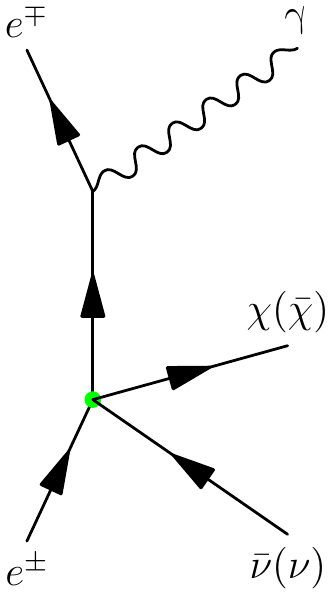}
}
\quad
\subfigure[]{
\label{fig:nca:Feyn:wwf}
\includegraphics[height=0.18\textheight]{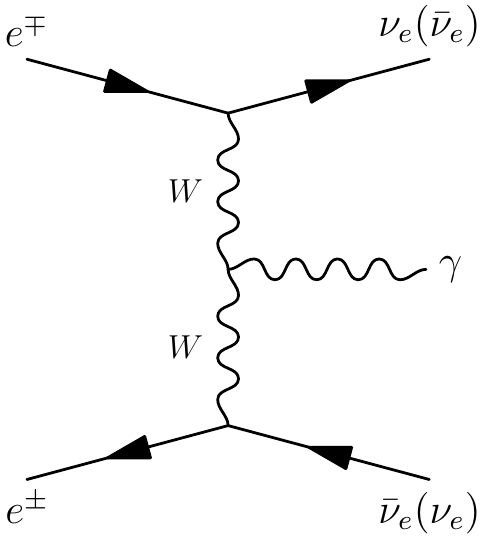}
}
\quad
\subfigure[]{
\label{fig:nca:Feyn:isrw}
\includegraphics[height=0.18\textheight]{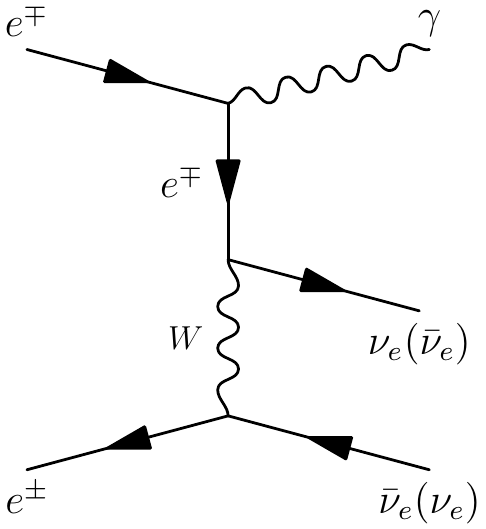}
}
\quad
\subfigure[]{
\label{fig:nca:Feyn:moz}
\includegraphics[height=0.18\textheight]{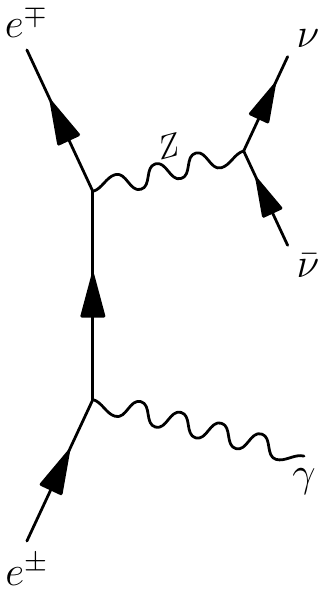}
}
\caption{\it The Feynman diagrams of the mono-photon production with
$e^+e^-$ collision, for signal (a) as well as the semi-irreducible
backgrounds from the $W$ mediated charged current (b)+(c)
and $Z$ resonant (d) channels.}
\label{fig:nca:Feyn}
\end{figure}

Observing the mono-photon radiation is
one of the most effective approach to search for dark particles at colliders
\cite{Fox:2011fx,Chae:2012bq,DELPHI:2003dlq,Birkedal:2004xn,Liu:2019ogn,Barman:2021hhg,Wan:2014rhl,Yu:2014ula,Dutta:2017ljq,Grzadkowski:2020frj,Ma:2022cto}. 
Missing energy and momentum are carried away by dark particles while the
photon energy is observable. 
If the invisible particle is a single bosonic dark particle, the
mono-photon is actually mono-energetic. Otherwise, the mono-photon
has continuous spectrum. We consider
possible test of a dark fermion
at lepton colliders in the associated production with
a neutrino. In contrast to the
DM pair $\chi \bar \chi$ production as usually
studied (for instance in \cite{Barman:2021hhg}), 
the effective operators in \geqn{eq:operators} will induce
events containing a single dark fermion $\chi$ (or $\bar \chi$)
and an associated anti-neutrino $\bar \nu$ (or a neutrino $\nu$),
as shown in \gfig{fig:nca:Feyn:isr}.
Both dark fermion $\chi$ and neutrino can carry missing energy and momentum.

For heavy dark fermion, the decay width of
$\chi\to\nu e^+e^-$ can be large. So it is possible for
the dark fermion be partialy visible.
Detecting the decay process can provide stronger
constraint with extra information.
To be conservative, we only consider an invisible
dark fermion with missing energy.

\subsection{Signal}

Two subprocesses, $e^+e^- \to \gamma\chi \bar{\nu}$ and
$e^+e^- \to \gamma \bar{\chi}\nu$ by exchanging a
$t$-channel electron in the first Feynman diagram
of \gfig{fig:nca:Feyn}, can contribute to the signal.
In the massless limit for the incoming $e^\pm$ beams, 
the differential polarized cross sections
$\sigma_{\lambda_{e^{-}}  \lambda_{e^{+}} }$
summed over the two subprocesses are,
\begin{subequations}
\begin{align}
  \frac{ d\sigma_{\pm\pm}^{S,P} }{ d m_{X}^{2} d\cos\theta_\gamma }
& =
  \frac{  1 }{32\pi^3 \Lambda^{4} }
  \frac{ (m_{X}^{2} - m_{\chi}^{2} )^2( s^2 + m_{X}^{4}) }{ s^2 ( s - m_{X}^{2}) m_{X}^{2} \sin^2\theta_\gamma } \,,
\label{eq:xs:sig:osp}
\\[3mm]
\label{eq:xs:sig:ova}
\frac{ d\sigma_{\pm\mp}^{V,A} }{ dm_{X}^{2} d\cos\theta_\gamma }
& =
\frac{ 1  }{384\pi^3 \Lambda^{4} }
\frac{ (m_{X}^{2} - m_{\chi}^{2} )^2( 2 m_{X}^{2} + m_{\chi}^{2} ) s }
{ ( s - m_{X}^{2}) m_{X}^{6} \sin^2\theta_\gamma }
\left( f_{V,A} + g_{V,A} \cos\theta_\gamma \right) \,,
\\[3mm]
\label{eq:xs:sig:ot}
\frac{ d\sigma_{\pm\pm}^{T} }{ dm_{X}^{2} d\cos\theta_\gamma }
& =
\frac{ 1  }{768\pi^3 \Lambda^{4} }
\frac{ (m_{X}^{2} - m_{\chi}^{2} )^2(  m_{X}^{2} +  2 m_{\chi}^{2}  ) s^2}{( s - m_{X}^{2}) m_{X}^{8}  \sin^2\theta_\gamma }
\left( f_{T} + g_{T} \cos\theta_\gamma \right) \,,
\end{align}
\label{eq:xs:sig}
\end{subequations}
where $\lambda_{e^\pm}=\pm1$ are the electron and positron 
helicities. Since neither the dark fermion $\chi$
nor the SM neutrino
$\nu$ are detectable at lepton colliders, it makes no
difference to treat the $\chi \nu$ system altogether as
a single particle with missing invariant mass 
$m_{X}^{2} \equiv (p_\nu + p_\chi)^2$.
For convenience, we have defined functions
$f_{V,A,T}$ and $g_{V,A,T}$, 
\begin{subequations}
\begin{align}
  f_{V,A}
& \equiv
  1 + 2 z_X + 10  z_X^2 + 2 z_X^3 + z_X^4,
\\
  g_{V,A}
& \equiv
  ( 1 - z_X )^2 (1 + 4 z_X + z_X^2 ),
\\
  f_{T}
& \equiv
  17 -4 z_X + 7z_X^2 + 56 z_X^3 + 55 z_X^4 - 4z_X^5 + z_X^6,
\\
  g_{T}
& \equiv
  (1- z_X)^3 (15 + 33 z_X + 15 z_X^2 + z_X^3),
\end{align}
\end{subequations}
in terms of $z_X \equiv m_X^2/s$. Besides $m_X$,
the only internal property of the $\chi \nu$ system
that appears is the dark fermion mass $m_\chi$ while the parameter
from the photon side is its scattering angle $\theta_\gamma$.
For the scalar and pseudo-scalar operators,
the situation is much simpler.

In the massless limit for electrons, which is an excellent
approximation at high energy colliders, the cross section
vanishes for the (pseudo-)scalar and tensor operators with
the helicity combinations $\lambda_{e^-} = - \lambda_{e^+}$,
and for the (axial-)vector operator with 
$\lambda_{e^-} =  \lambda_{e^+}$. 
Since the associated photon in \gfig{fig:nca:Feyn:isr}
arises from the initial state radiation (ISR),
the cross sections exhibit a singular behavior 
at the colinear limit $\sin\theta_\gamma = 0$ which
explains the $1/\sin^2 \theta_\gamma$ factor in
\geqn{eq:xs:sig}. Although this divergent behavior
can be regularized by the electron mass, it captures
the correct feature in the forward and backward regions.
Furthermore, signals are also singular
if the emitted photon is very soft
(when $m_{X}^2 \to s$) as summarized by the factor 
$1/(s - m_{X}^{2})$. Considering this, our numerical
calculations adopts the following kinematic
cuts to enhance the simulation efficiency
\cite{Habermehl:2020njb},
\begin{equation}
  4^\circ < \theta_{\gamma} < 176^\circ
\,\,\,
  ( |\eta_\gamma| < 3.35 ),
\quad \mbox{and} \quad
  p_{T,\gamma} > 1\gev.
\label{eq:gen:cuts}
\end{equation}
Our numerical simulations are conducted in 
\texttt{MadGraph5} \cite{Alwall:2014hca} with model
files generated by
\texttt{FeynRules} \cite{Alloul:2013bka}.

\begin{figure}[t]
\centering
\includegraphics[width=0.48\textwidth]{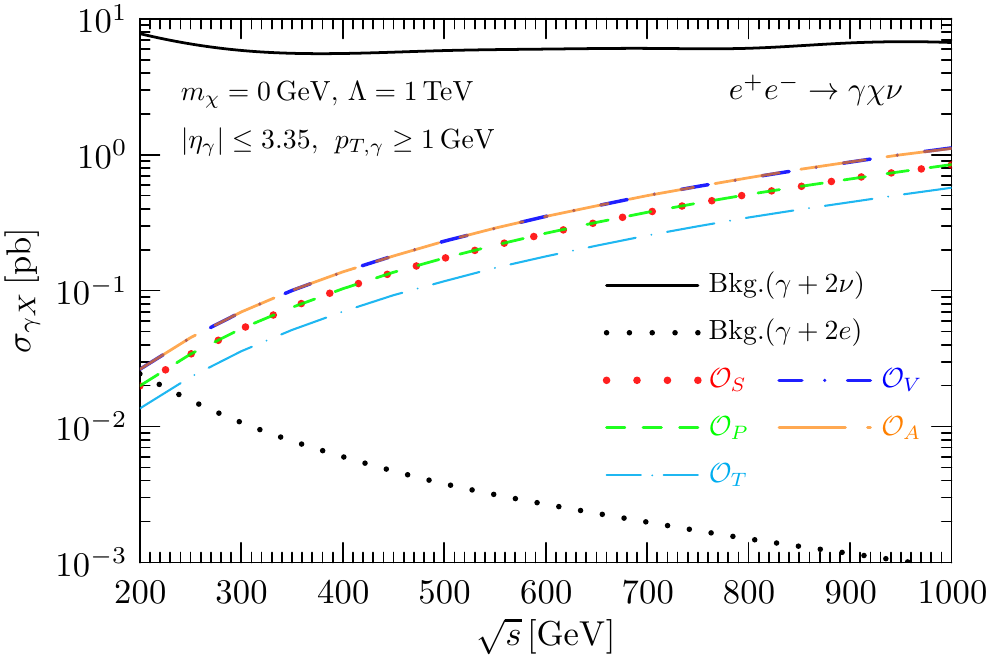}
\quad
\includegraphics[width=0.48\textwidth]{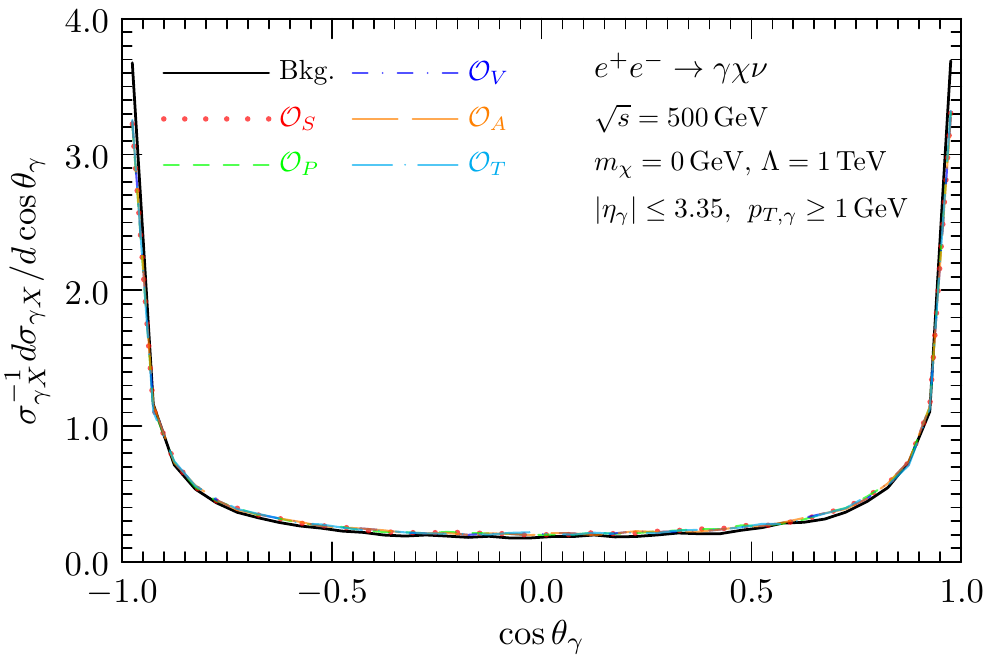}
\caption{\it 
{\bf Left panel}: the mono-photon production cross sections 
as function of the collision energy $\sqrt{s}$. The signal ones
(colorful non-solid curves) are obtained with vanishing
dark fermion mass $m_{\chi}=0\gev$ and cut-off scale $\varLambda_i=1\tev$.
The background includes the semi-irreducible
$e^+e^- \to \gamma + 2\nu$ (black-solid line) and the
reducible one from the radiative Bhabha scattering
$e^+e^- \to e^+e^- + 2\nu$ (black-dotted line). 
{\bf Right panel}: the normalized photon polar angle
($\theta_\gamma$) distributions in the lab frame with
collider energy $\sqrt{s}=500\gev$. The kinematic
cuts $p_{T,\gamma} < 1\gev$ and $|\eta_{\gamma}| < 3.35$
are in place.}
\label{fig:xe-yxsp}
\end{figure}

The left panel of \gfig{fig:xe-yxsp} shows the
collision energy $\sqrt{s}$ dependence
of the total cross sections with the kinematic
cuts given in \geqn{eq:gen:cuts}. For illustration,
signals are simulated with massless DM particle
($m_{\chi}=0\gev$) and $\varLambda=1\tev$. One
can see that all the signal cross sections 
grow with $\sqrt{s}$.
Contributions of the vector (blue dash-dotted) 
and axial vector (yellow long-dashed) operators 
are slightly larger than the scalar
(red dotted) and pseudo-scalar (green dashed) 
ones while the tensor operator
(cyan long-dash-dotted) has the smallest cross section.
It is interesting to observe that there is no
difference between the vector and axial-vector
or between the scalar and pseudo-scalar operators.
The right panel shows the normalized distribution as
function of the photon polar angle $\theta_\gamma$
in the lab frame with a collision energy
$\sqrt{s} = 500\gev$.
As expected, for all the five effective operators,
signals are significantly populated
in the forward and backward regions with nearly
indistinguishable polar angle distributions.

\subsection{Background}

The dominant background is the radiative
neutrino pair production, \ie, neutrino pair production
together with a photon, as shown in \gfig{fig:nca:Feyn:wwf},
\gfig{fig:nca:Feyn:isrw}, and \gfig{fig:nca:Feyn:moz}.
Although the background does not share exactly the same
final-state fermions with signal, neither neutrino nor
dark matter is detectable at collider and hence we
call such background as semi-irreducible.
Note that other background diagrams with the SM Higgs
boson are negligible since the electron Yuakwa coupling
is quite tiny.
From the left panel of \gfig{fig:xe-yxsp}, one can see that
at $\sqrt{s}=1\tev$, the signals are about 1 order smaller
than the semi-irreducible background (black-solid curve).
Furthermore, the associated photon is emitted via
ISR and hence the
corresponding amplitude is singular when approaching
$\sin\theta_\gamma = 0$. Consequently, both signals
and the semi-irreducible background favor the forward and
backward regions, as shown in the right panel of
\gfig{fig:xe-yxsp}.

Since photon is the only detectable particle for the
signal, a mono-$\gamma$ production with any other
undetected particles can also contribute to the total
background. The major contribution comes from the
radiative Bhabha scattering ($e^+e^- \to \gamma e^+e^-$)
of which the final-state $e^+ e^-$ evade detection
in the very forward region \cite{Bartels:2012ex}.
Such Bhabha scattering has huge
cross section and hence can potentially give a sizable
contribution as reducible background.
Note that the divergent configurations for the
radiative Bhabha scattering should be removed at
the generator level
\cite{Habermehl:2020njb,Berends:1983fs,Tobimatsu:1985vd}
following the treatments in \cite{Habermehl:2018yul},
\begin{subequations}
\begin{align}
&
m(e^{\pm}_{\rm in}, e^{\pm}_{\rm out}) < -1\gev \,, \quad
m(e^{-}_{\rm out}, e^{+}_{\rm out}) > 1\gev\,, \quad
m(e^{\pm}_{\rm out}, \gamma) > 4\gev\,, 
\\
&
0.1\gev < p_{T, e^{\pm}_{\rm out} } < 1\gev \,, \quad
\Delta R(e^{-}_{\rm out}, e^{+}_{\rm out}) > 0.4 \,, \quad
\Delta R(e^{\pm}_{\rm out}, \gamma) > 0.4 \,.
\end{align}
\end{subequations}
This Bhabha scattering background can
be further suppressed to the per mille level by vetos
based on reconstructed objects in the electromagnetic
calorimeters in the very forward region (BeamCal)
\cite{Habermehl:2020njb} by a factor of $0.23\%$
\cite{Habermehl:2020njb,Habermehl:2018yul}.
The black-dotted curve in the left panel of
\gfig{fig:xe-yxsp} shows the 
radiative Bhabha scattering contribution to the total
background. One can see that it is more than 2 orders
of magnitude smaller than the semi-irreducible one 
and hence is safely negligible. Unless otherwise
specified, our discussion focuses on the
semi-irreducible background.

For $\sqrt s \leq 1$\,TeV, the SM background dominates
over the possible signal with cut-off scale
$\Lambda = 1$\,TeV by at least one order. The situation
becomes even worse for a massive DM since a nonzero
$m_\chi$ will further reduce the final-state phase
space. To detect
such signal or put meaningful limit on the
absorption operators, energy cuts are necessary to
suppress background and enhance the signal significance.

\subsection{Energy Cut}

\begin{figure}[t]
\centering
\includegraphics[width=0.7\textwidth]{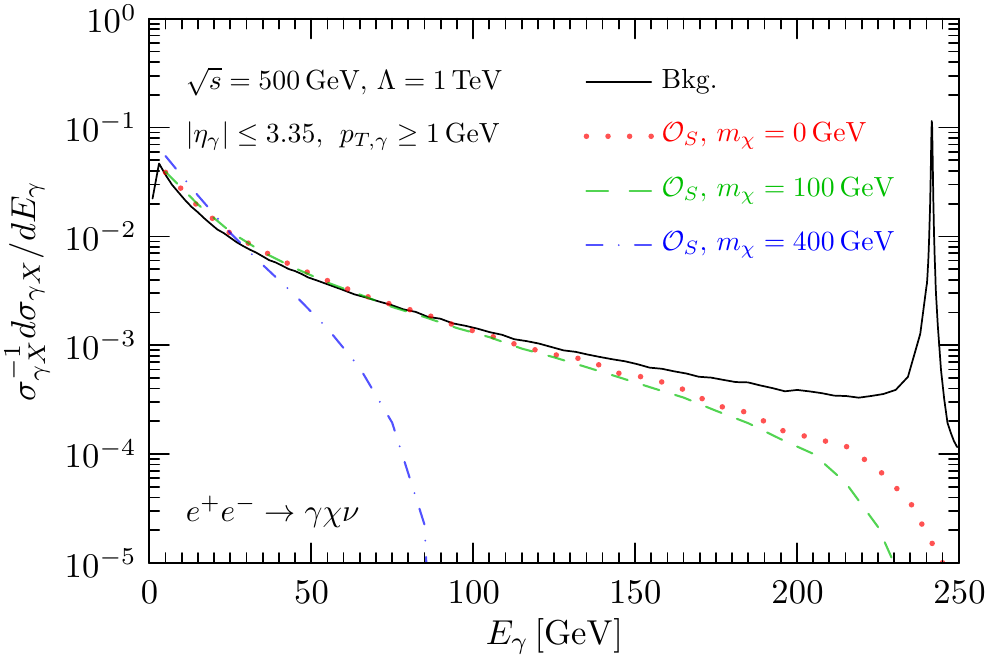}
\caption{\it 
The normalized photon energy $E_\gamma$ distribution in
the lab frame with $\sqrt{s} = 500\gev$. For illustration,
signals are shown with several dark fermion masses, $m_{\chi}=0\gev$
(red-dotted), $100\gev$ (green-dashed), and $400\gev$
(blue-dash-dotted). The same kinematic cuts
$p_{T,\gamma} < 1\gev$ and $|\eta_{\gamma}| < 3.35$ have
been used as before.}
\label{fig:xe-yxs}
\end{figure}

Since the photon polar angle $\theta_\gamma$ cannot
be used to distinguish signals and background, it is
desirable to find another observable for the purpose
of enhancing the signal-background ratio. With photon
being the only detectable particle, one has to resort
to the photon property. In addition to $\theta_\gamma$,
the only observable is the photon energy. 

\gfig{fig:xe-yxs} shows the photon energy ($E_\gamma$)
spectrum at collision energy $\sqrt{s} = 500\gev$. We can
clearly see a peak in the background spectrum (black
solid) which is due to the $Z$ resonance in
\gfig{fig:nca:Feyn:moz}. For a two-body phase space
of the $Z \gamma$ system, the photon energy is
determined by the $Z$ boson mass $m_Z$ and the
collision energy $\sqrt s$,
\bee
  E^Z_\gamma
=
  \frac 1 2 \sqrt{s}
\left( 1 - \frac{ m_Z^2 }{s} \right).
\ene
The peak width is mainly determined by the $Z$ boson
width and the ISR. For comparison, the signals (colorful
curves) do
not have a peak at that place but a continuously
decreasing photon spectrum that finally stops at,
\bee
  E^\chi_{\gamma}
\le
  \frac{1}{2}\sqrt{s} \left( 1 - \frac{ m_\chi^2 }{s} \right).
\label{eq:Echi}
\ene
This upper limit is reached when the $\chi \bar \nu$
($\bar \chi \nu$) system has the minimal invariant mass,
$m_X \geq m_\chi$.

\begin{figure}[t]
\centering
\includegraphics[width=0.7\textwidth]{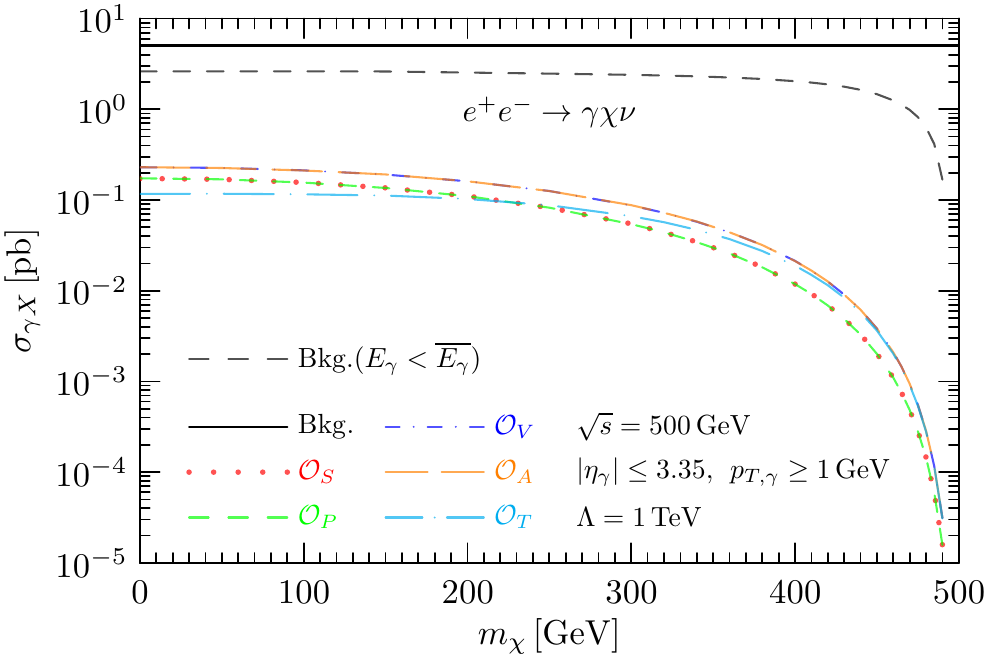}
\caption{\it 
The mono-photon production cross sections at $\sqrt{s}=500\gev$
as function of the dark fermion mass $m_{\chi}$.}
\label{fig:Xsec-mChi}
\end{figure}

With such difference in the photon energy spectrum,
it is possible to cut off a significant part of the
background with
\bee
  E_\gamma < \overline E_\gamma
\equiv
  \min\left\{ E_\gamma^{Z} - 5 \varGamma_{Z}, E_\gamma^{\chi} \right\},
\label{eq:EgammaCut}
\ene
to enhance the signal significance.
The black-dashed curve in \gfig{fig:Xsec-mChi} shows the
irreducible background after applying the above
kinematic cut. For small $m_\chi$, the background
event rate reduces by roughly a factor of 2. Although
the $Z$ resonance peak can be removed, the low-energy
peak with $E_\gamma$ approaching 0, which is caused
by the $W$ mediated diagrams in \gfig{fig:nca:Feyn:isrw},
almost overlaps with the signal spectrum and hence
cannot be effectively suppressed. It is interesting
to observe that larger reduction can be achieved for
larger dark fermion mass $m_\chi$ as shown by the black dashed
line in \gfig{fig:Xsec-mChi}. This is because the
photon energy cut $\overline E_\gamma$ is a function
of the dark fermion mass $m_\chi$
via the $E^\chi_\gamma$ defined in \geqn{eq:Echi}.

\gfig{fig:Xsec-mChi} shows the $m_\chi$ dependence
of the total cross sections with the kinematic cuts
in \geqn{eq:gen:cuts} and \geqn{eq:EgammaCut}.
As expected, the signal cross sections decrease with
the dark fermion mass $m_{\chi}$ and vanish when approaching 
the kinematically allowed upper limit,
$m_{\chi} \leq \sqrt{s}$. Even after applying the
energy cut, the SM background still dominates over
signals with $\Lambda = 1$\,TeV. More
powerful measures such as the beam polarizations
are necessary as we elaborate below.

\subsection{Beam Polarization}
\label{sec:beamA}

After suppressing the
$Z$ resonance peak by photon energy cut in
\geqn{eq:EgammaCut}, the major SM background
comes from the $W$ mediated diagrams in
\gfig{fig:nca:Feyn}. Note that the charged current
interactions mediated by the SM $W$ only involve
left-handed fermions. This allows the possibility
of using the beam polarizations for further
suppression of background which is one of the
most important advantages of $e^+e^-$ colliders.

The cross sections with electron and positron beam
polarization rates $P_{e^-}$ and $P_{e^+}$, respectively,
are given (in helicity basis) by,
\bea
  \sigma\big(P_{e^-}, P_{e^+}\big)
=
  \sum_{ \lambda_{e^{-}} , \lambda_{e^{+}} = \pm 1}
  \frac {1 + \lambda_{e^{-}} P_{e^-}} 2
  \frac {1 + \lambda_{e^{+}} P_{e^+}} 2
  \sigma_{\lambda_{e^{-}}  \lambda_{e^{+}}},
\label{eq:sigmaP}
\ena
where $\sigma_{\lambda_{e^{-}} \lambda_{e^{+}} }$,
are the cross sections with purely polarized beams,
$\lambda_{e^\pm} = \pm 1$, in the
helicity basis. The electron and positron polarization
rates are defined
as $P_{e^\pm} \equiv (\phi_{e^\pm_\uparrow} - \phi_{e^\pm_\downarrow}) / (\phi_{e^\pm_\uparrow} + \phi_{e^\pm_\downarrow})$
with $\phi_{e^\pm_{\uparrow,\downarrow}}$ being
the positron/electron fluxes 
with positive ($\uparrow$) and negative ($\downarrow$)
helicities, respectively. Then the prefactor
$(1 + P_{e^\pm}) / 2$ is the weight for each
beam polarization component.

\begin{figure}[t]
\centering
\includegraphics[width=0.48\textwidth]{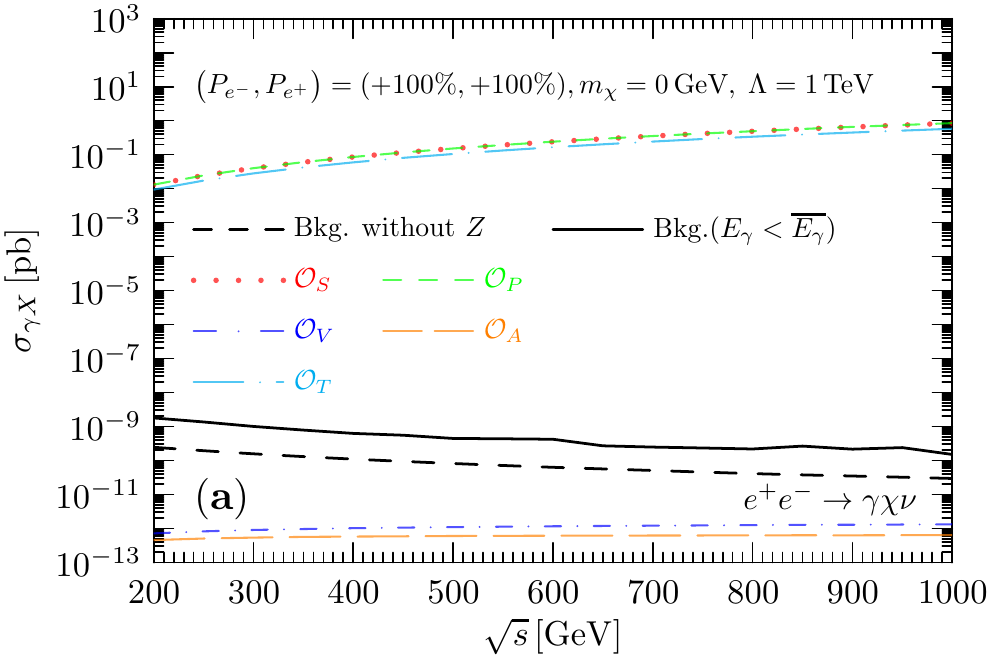}
\hfill
\includegraphics[width=0.48\textwidth]{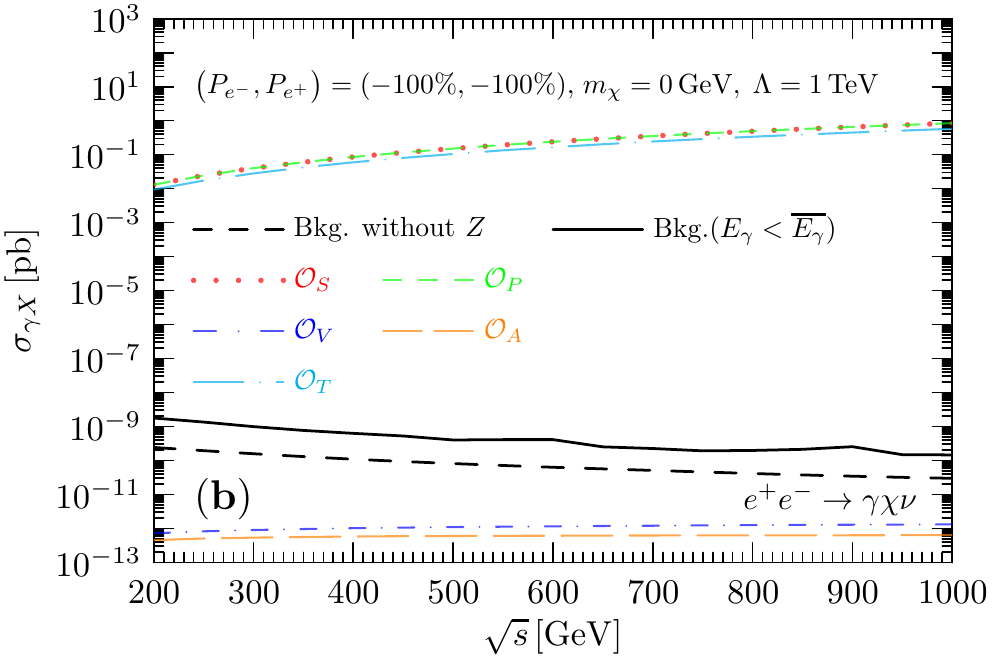}
\\[2mm]
\includegraphics[width=0.48\textwidth]{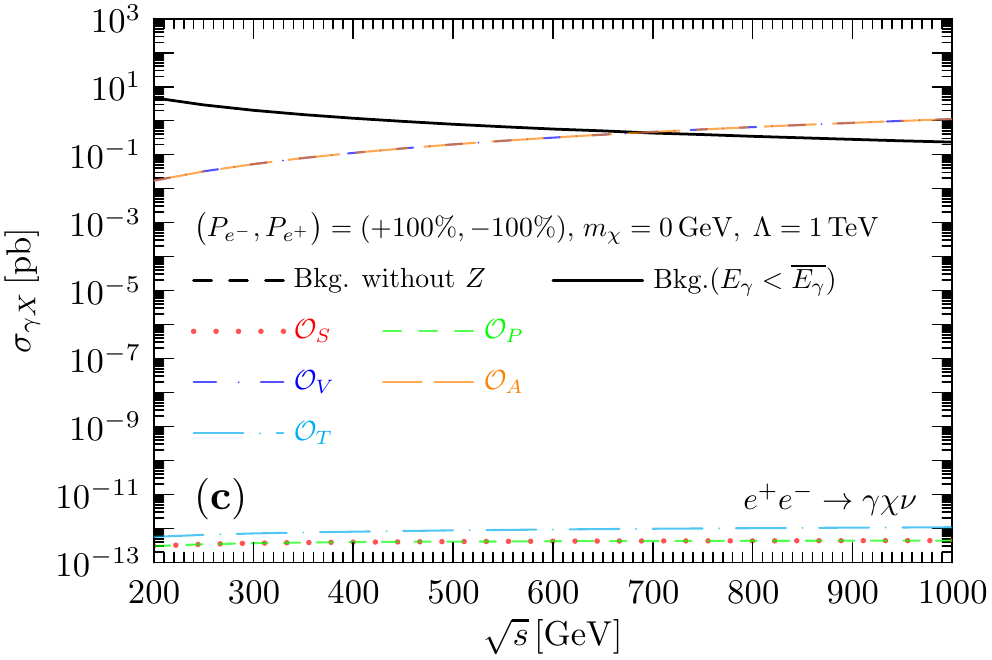}
\hfill
\includegraphics[width=0.48\textwidth]{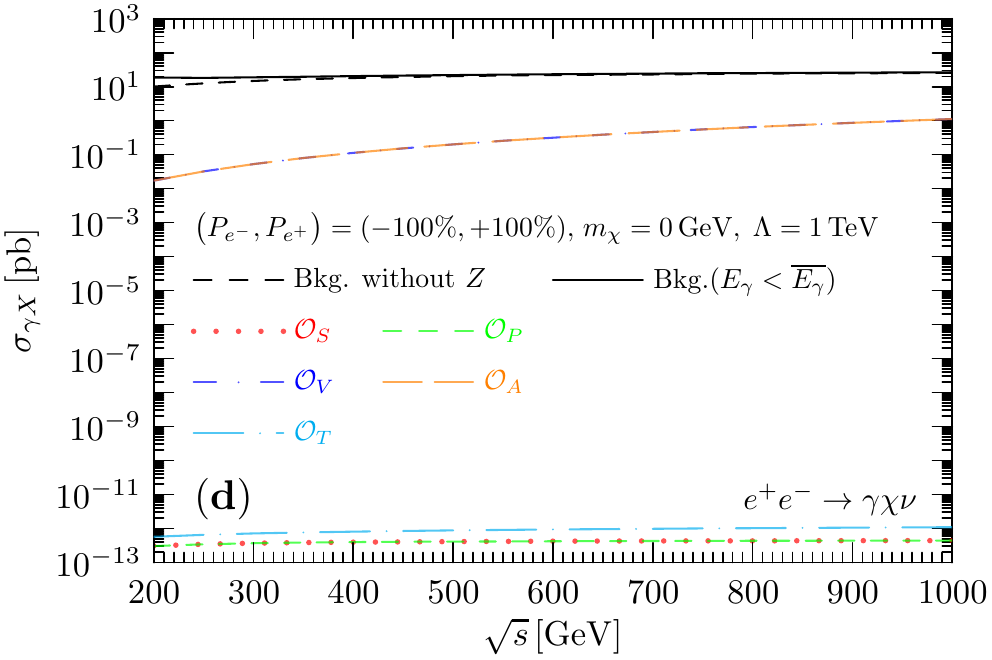}
\caption{\it The mono-photon cross section with purely polarized
beams $(P_{e^-},\, P_{e^+})=(+100\%, +100\%)$ (top-left),
$(-100\%, -100\%)$ (top-right), $(+100\%, -100\%)$ (bottom-left),
and $(-100\%, +100\%)$ (bottom-right), respectively.
The signal ones (colorful non-solid) are shown with
vanishing dark fermion mass $m_{\chi}=0\gev$ and cut-off scale
$\varLambda_i=1\tev$. The black curves show the semi-irreducible
backgrounds with (black solid) and without (black dashed)
the $Z$ resonance contribution. Note that the black-dashed curve
with beam polarization $(P_{e^-},\, P_{e^+})=(+100\%, -100\%)$
is too small ($\sim 10^{-24}$\,{\rm pb} ) and lies
outside the plot region.}
\label{fig:xe-yxs:eipi}
\end{figure}

\gfig{fig:xe-yxs:eipi} shows the polarization effects
on the signal (colorful non-solid lines) and background (black)
cross sections. The electron and positron beams have
$P_{e^\pm} = \pm 100\%$ polarizations. Since only the
left-handed electron and positron can participate the
SM charged current interactions, either
$P_{e^-} = +100\%$ or $P_{e^+} = - 100\%$ suppresses
the two Feyman diagrams in the middle of
\gfig{fig:nca:Feyn} explaining why the black dashed
lines in the first three subplots of
\gfig{fig:xe-yxs:eipi} are highly suppressed. To make
this effect more transparent, the black dashed
lines are obtained with the $Z$ resonance being
manually removed and only the $W$ mediated diagrams
included. In particular, \gfig{fig:xe-yxs:eipi}(c)
even does not contain a black dashed line due to
the double suppression of both $P_{e^-} = +100\%$
and $P_{e^+} = - 100\%$. The residual contributions
for the two black dashed lines in \gfig{fig:xe-yxs:eipi}(a)
and \gfig{fig:xe-yxs:eipi}(b) exist only because
of the chirality flip with a nonzero electron mass.
For $\sqrt s = 500$\,GeV, the size of the chirality
flip effect is roughly
$m_e / \sqrt s \sim \mathcal O(10^{-6})$ at the amplitude
level which is squared to account for the $\mathcal O(10^{-12})$
suppression factor at the cross section level.
With double suppression for \gfig{fig:xe-yxs:eipi}(c),
the black dashed line with $\mathcal O(10^{-24})$\,pb
is way out of the plot range. For the polarization
combinations $P_{e^-} = -100\%$ and $P_{e^+} = +100\%$
in \gfig{fig:xe-yxs:eipi}(d), the $W$ contribution
is almost not affected. To suppress the $W$ mediated
contributions, it is better to have either
$P_{e^-} = +100\%$ or $P_{e^+} = -100\%$.

For the $Z$ resonance, the electron and positron
polarizations need to be chosen coherently since
no single one can suppress its contribution. This
is because the $Z$ boson mediated neutral current
can couple with both left- and right-handed fermions.
However, the vector and axial-vector currents do
not flip chirality. With mismatched polarizations,
either $P_{e^-} = +100\%$ with $P_{e^+} = +100\%$
or $P_{e^-} = -100\%$ with $P_{e^+} = -100\%$, the
$Z$ resonance can also be suppressed. These can be
seen in the two suppressed black solid lines in the
first two panels of \gfig{fig:xe-yxs:eipi}.

Combing the requirements for suppressing the $W$
and $Z$ mediated diagrams, it seems only the two
combinations in the first two panels of
\gfig{fig:xe-yxs:eipi} satisfy our purpose.
Unfortunately, the signals will also be affected,
especially for the vector and axial-vector type
operators in \geqn{eq:operators}. With the same
chirality structure as the SM neutral current
interactions, the cross section for these two
operators will also be significantly suppressed.
Non-pure beam polarization also has its reason
to stay.

\begin{figure}[t]
\centering
\includegraphics[width=0.48\textwidth]{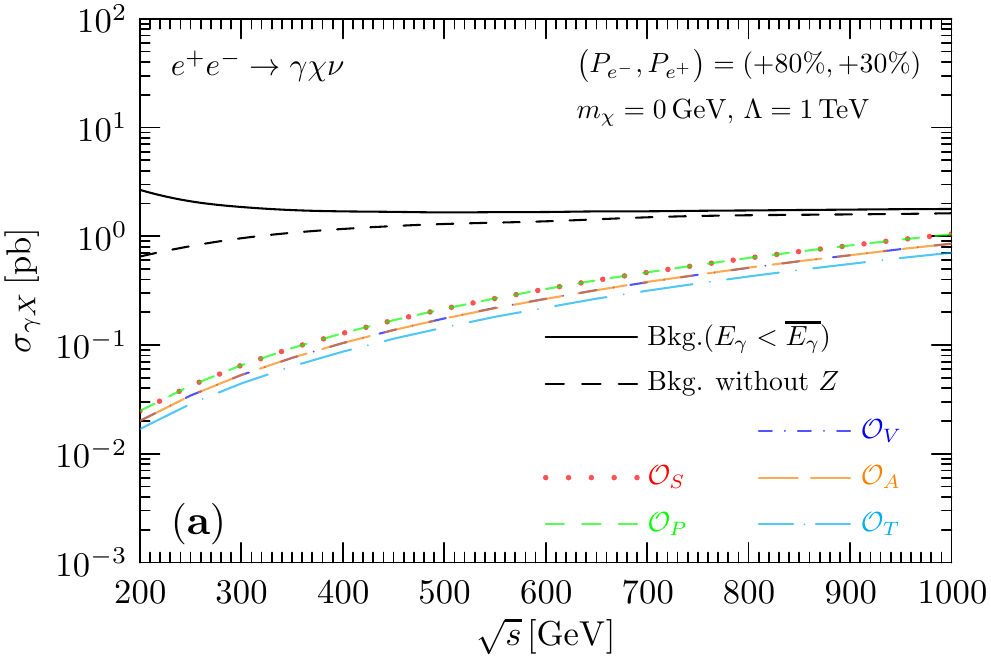}
\quad
\includegraphics[width=0.48\textwidth]{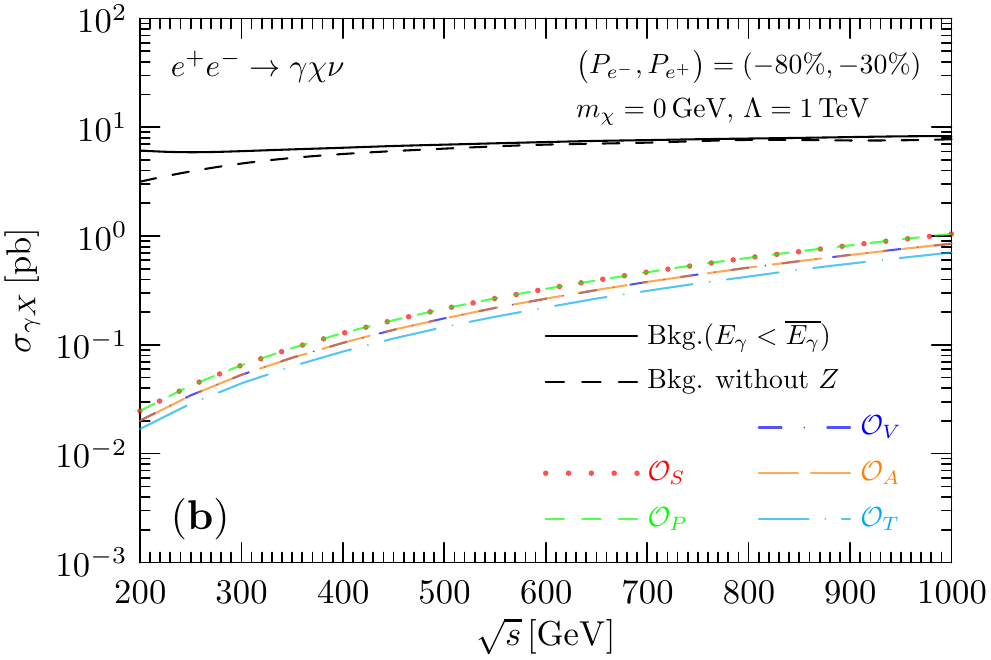}
\\[2mm]
\includegraphics[width=0.48\textwidth]{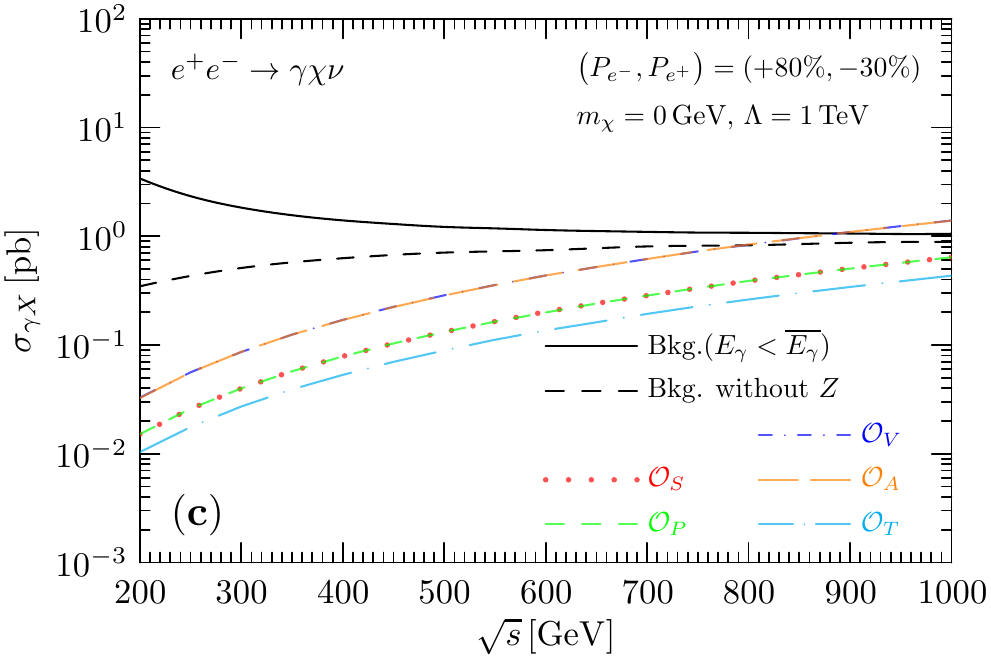}
\quad
\includegraphics[width=0.48\textwidth]{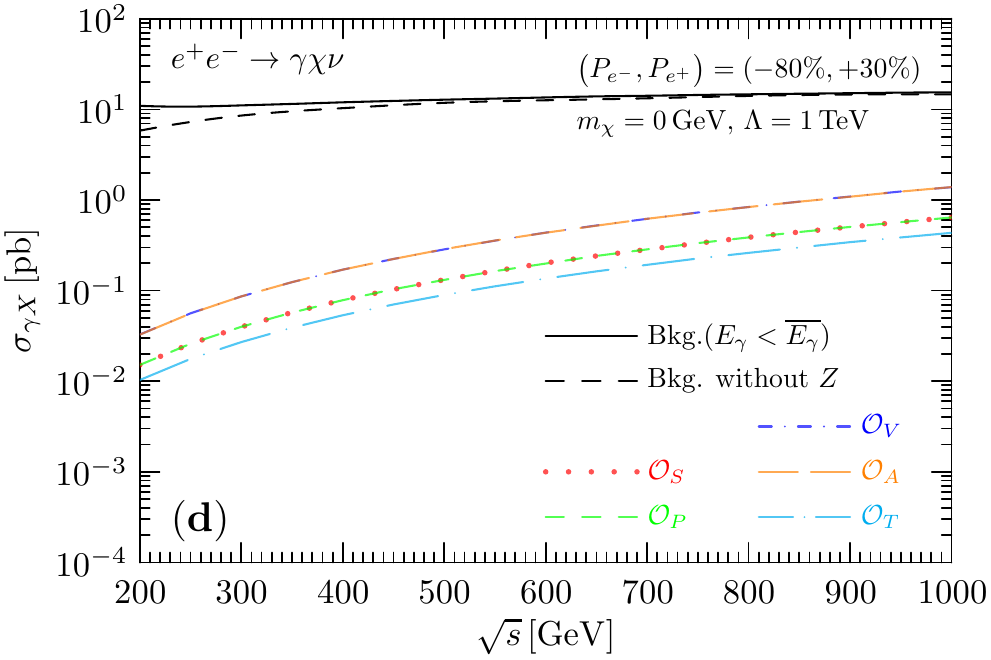}
\caption{\it
The mono-photon cross sections with impure beam polarizations
$(P_{e^-},\, P_{e^+})=(+80\%, +30\%)$ (top-left),
$(-80\%, -30\%)$ (top-right), $(+80\%, -30\%)$ (bottom-left),
and $(-80\%, +30\%)$ (bottom-right), respectively.}
\label{fig:xe-yxs:e80p30}
\end{figure}

Another reason for impure polarization is the practical
possibility. It is very difficult to make a beam 100\%
polarized. \gfig{fig:xe-yxs:e80p30} shows the polarized
cross sections with typical beam polarizations
$P_{e^-} = \pm 80\%$ and $P_{e^+} = \pm 30\%$. 
Again, the black solid (dashed) curves show the result
with (without) the $Z$ resonance contribution. For the
$W$ only result (black dashed line), the only sizable
contribution comes from the polarization configuration
of $P_{e^-} = -100\%$ and $P_{e^+} = +100\%$ in
\gfig{fig:xe-yxs:eipi}(d). Consequently, the black dashed
lines in all the panels of \gfig{fig:xe-yxs:e80p30}
can be simply rescaled according to the polarization
weighting factors in \geqn{eq:sigmaP}. The largest
one is $\sigma_{\rm Bkg}(-80\%, +30\%)$ in
\gfig{fig:xe-yxs:e80p30}(d) which is approximately
58.5\% of the maximally polarized one
$\sigma_{\rm Bkg}(-100\%, +100\%)$. The other channels
are roughly suppressed by a factor 
of $3.5\%$ for $\sigma_{\rm Bkg}(+80\%, -30\%)$
(bottom-left panel), $6.5\%$ for
$\sigma_{\rm Bkg}(+80\%, +30\%)$ (top-left panel),
and $31.5\%$ for $\sigma_{\rm Bkg}(-80\%, -30\%)$
(top-right panel). Although the $W$ mediated contributions
are suppressed to be negligibly small in the first
three panels of \gfig{fig:xe-yxs:eipi}, its effect
remains quite sizable for impure polarization
configurations. The similar thing also happens for
the $Z$ resonance. In all the four panels of
\gfig{fig:xe-yxs:e80p30}, the SM background almost
always dominates over signals. Nevertheless, the
energy cut and polarization already bring the SM
background event rate quite close to the signal ones.
With the expected high precision measurement, it is
still quite promising to probe the fermionic
absorption DM scenario ($m_\chi = 0$\,GeV)
at the future lepton colliders.

Impure polarization also has a benefit of allowing
a sizable event rate for all types of effective operators.
The signals (colorful non-solid curves) in
\gfig{fig:xe-yxs:e80p30} are shown with
vanishing dark fermion mass $m_{\chi} = 0\gev$ and cut-off scale
$\varLambda_i=1\tev$. We can see that for the five
operators, their cross sections are quite close to
each other. The smallest difference between the signal
and background cross sections happens for the left
two panels with $P_{e^-} = 80\%$ and particularly
together with $P_{e^+} = -30\%$ which can be treated
as our optimal choice or requirement.

\subsection{Projected Sensitivity}

There are several proposals for the future $e^{+}e^{-}$
colliders, CEPC \cite{CEPCStudyGroup:2018ghi},
FCC-ee \cite{FCC:2018evy,FCC:2018byv}, ILC
\cite{Bambade:2019fyw,Barklow:2015tja}, and
CLIC \cite{Aicheler:2018arh}. While circular
colliders (CEPC and FCC-ee) can provide higher
luminosity, the linear ones (ILC and CLIC)
can reach higher energy and are much easier to
have polarized beams.
\begin{itemize}
\item 
The CEPC has three different running modes
\cite{CEPCStudyGroup:2018ghi}, the Higgs factory mode
at $\sqrt{s} = 240\gev$ with a total luminosity of
$5.6\ab^{-1}$, the $Z$ factory mode at $\sqrt{s} =
91.2\gev$ with a total luminosity of $16\ab^{-1}$, and
the $WW$ threshold scan in the range of
$\sqrt{s} \sim 158 - 172\gev$ with a total 
luminosity of $2.6\ab^{-1}$. As illustrated earlier,
the signal cross sections grow with $\sqrt{s}$.
So in this paper we consider only the Higgs factory
mode that has the highest collision energy at CEPC.
Although the beam polarization is difficult at circular
colliders, there are already some explorations
\cite{Duan:2019szc,Duan:2023lyp}.

\item 
The FCC-ee also runs at the $Z$ pole with a projected
luminosity of 150\,ab$^{-1}$ and the Higgs mode
($\sqrt{s}=240$\,GeV) with a total luminosity of 5\,ab$^{-1}$
\cite{FCC:2018evy,FCC:2018byv}. 
The Running mode near the $W$-pair production threshold,
$\sqrt{s}\sim 162.5$\,GeV, can accumulate a total luminosity
of 12\,ab$^{-1}$. Around the top-pair production threshold,
there is a multi-point scan at eight collision energies $\sqrt s =$
(340, 341, 341.5, 342, 342.5, 343, 343.5, 344, and 345)\,GeV
each with a luminosity of $25$\,fb$^{-1}$.
The beam polarization has also been proposed which
is under study for the $Z$-pole and $W$-pair
production modes
\cite{Mane:2014nqa,Blondel:2019jmp,Blondel:2021zix}.
Although the top-pair production mode has much higher
energy, the luminosity is much smaller. So the preferred
running mode at FCC-ee is still the Higgs factory one
with both high enough energy and large luminosity. Due to
similarity with CEPC for the Higgs factory mode, we take
CEPC as an example in later discussions and the results
can scale accordingly for FCC-ee.

\item
The ILC was proposed to run at $\sqrt{s}=500\gev$, and will collect a data
set of $4\ab^{-1}$ in total \cite{Bambade:2019fyw}. 
Furthermore, the beam polarization will be studied according to 
the H20 running scenario \cite{Barklow:2015tja} for optimizing the physics performance.
More recently, ILC running at $\sqrt{s}=250\gev$ with a total luminosity $2\ab^{-1}$ 
has also been proposed \cite{Fujii:2017vwa}.
Since it is much similar to the Higgs factory mode at CEPC, here we consider
only the case of $\sqrt{s}=500\gev$ which has much higher energy.
The ILC also plans to run at $\sqrt{s}=90$\,GeV and 160\,GeV
with smaller total luminosities of 100\,fb$^{-1}$ and 500\,fb$^{-1}$ \cite{Barklow:2015tja},
respectively.

\item
The CLIC is designed to run at 380\,GeV, 1.5\,TeV, and
3\,TeV with total integrated luminosities of 1\,ab$^{-1}$,
2.5\,ab$^{-1}$, and 5\,ab$^{-1}$, respectively
\cite{Aicheler:2018arh}. We focus on the highest energy
case of 3\,TeV with the largest luminosity for demonstration.
This does not mean the 1.5\,TeV mode is not important.
Even though running modes with an 80\% polarization
has also been proposed, it only applies for the electron
beam but not the positron one. So we only consider the
unpolarized case at CLIC.

\end{itemize}

\begin{table}[htp]
\centering
\begin{tabular}{c|c|c|c|c|c}
& CEPC & \multicolumn{3}{c|}{ILC}  & CLIC  
\\ \hline
$\sqrt{s}\,[\gev]$ & 240 & \multicolumn{3}{c|}{500} & 3000
\\ \hline
$(P_{e^-}, P_{e^+})$ & (0\%,\,0\%) & (0\%,\,0\%) & $(\pm80\%,\,\mp30\%)$
& $(\pm80\%,\,\pm30\%)$ & (0\%,\,0\%)  
\\ \hline 
$\call\,[\ab^{-1}]$ & 5.6 & 4 & 1.6 & 0.4 & 5
\end{tabular}
\caption{\it The running modes and the corresponding
projected luminosities at the future $e^+ e^-$
colliders.}
\label{tab:mechs}
\end{table}%
The typical running modes and projected luminosities
used in our simulations are
listed in \gtab{tab:mechs}.

As demonstrated earlier, the photon energy $E_\gamma$
is the only sensitive observable for distinguishing
signals and background. In practical measurement,
the photon energy is smeared by various factors such
as the ISR and detector resolutions
\cite{Habermehl:2020njb}. Since both signals
and background have continuous $E_\gamma$ distributions,
the spectrum is not significantly modified. But the
total cross sections are reduced with relatively
stronger dependence on the collision energy $\sqrt{s}$
that is depleted by the ISR.
In this paper, the ISR effect is taken into account
by using the plugin \texttt{MGISR}
\cite{Chen:2017ipx,Li:2018qnh} to \texttt{MadGraph5} \cite{Alwall:2014hca}.

The events at these future $e^+ e^-$ colliders are first
selected with the following minimal cuts,
\begin{subequations}
\bea
p_{T, \gamma} > 0.5\gev, \quad \big|\eta_\gamma \big| < 2.65\,,\quad\quad&&  \text{CEPC};
\\[2mm]
p_{T, \gamma} > 6\gev, \quad \big|\eta_\gamma \big| < 2.79\,,\quad\quad &&  \text{ILC};
\\[2mm]
p_{T, \gamma} > 60\gev, \quad \big|\eta_\gamma \big| < 2.44\,,\quad\quad &&  \text{CLIC}.
\ena
\end{subequations}
For CEPC, the rapidity cut is chosen according to
its Conceptual Design Report (CDR)
\cite{CEPCStudyGroup:2018ghi} for the
Electromagnetic Calorimeter (EMC) coverage
and the $p_{T,\gamma}$ cut is slightly larger to reduce
the radiative Bhabha background \cite{Liu:2019ogn}.
Due to the BeamCal configuration, two different cuts
$p_{T,\gamma} > 1.92\gev$ and $p_{T,\gamma} > 5.65\gev$
are needed at ILC
\cite{Habermehl:2020njb,Habermehl:2018yul}.
For simplicity, we use a universal transverse momentum cut
$p_{T,\gamma} > 6$\,GeV while the rapidity cut is taken
from \cite{Habermehl:2020njb}. The cuts for CLIC follow
those in \cite{Blaising:2021vhh}. 

The signal significance is also be affected by the photon
reconstruction efficiency. It can reach above 99\% for
$E_\gamma < 2\gev$ and $7^\circ < \theta_\gamma < 173^\circ$ 
($\big|\eta_\gamma \big| < 2.79$) at ILC
\cite{Habermehl:2018yul} and similarly at CLIC
\cite{Blaising:2021vhh}.
At CEPC, the identification efficiency is nearly 100\%
for photons with $E_\gamma > 5$\,GeV and
more than 99\% of their energy can be 
reconstructed~\cite{CEPCStudyGroup:2018ghi}.
Hence we simply take 100\% reconstruction efficiency
for illustration.

Since the photon energy $E_\gamma$ spectrum for signals
and background are quite similar in the signal dominating
region with small $E_\gamma$ (see \gfig{fig:xe-yxs}), the 
signal significance can not be enhanced much by data binning.
So we simply use the total event number to estimate
the experimental sensitivity with the following $\chi^2$,
\bee
  \chi^2
=
  \frac{ \big( N^{\rm Sig} \big)^2 }
{  N^{\rm Bkg} + N^{\rm Sig} } \,,
\ene
with $N^{\rm Bkg}$ and $N^{\rm Sig}$ being the background
and signal event numbers, respectively.

\begin{figure}[t]
\centering
\includegraphics[width=0.32\textwidth]{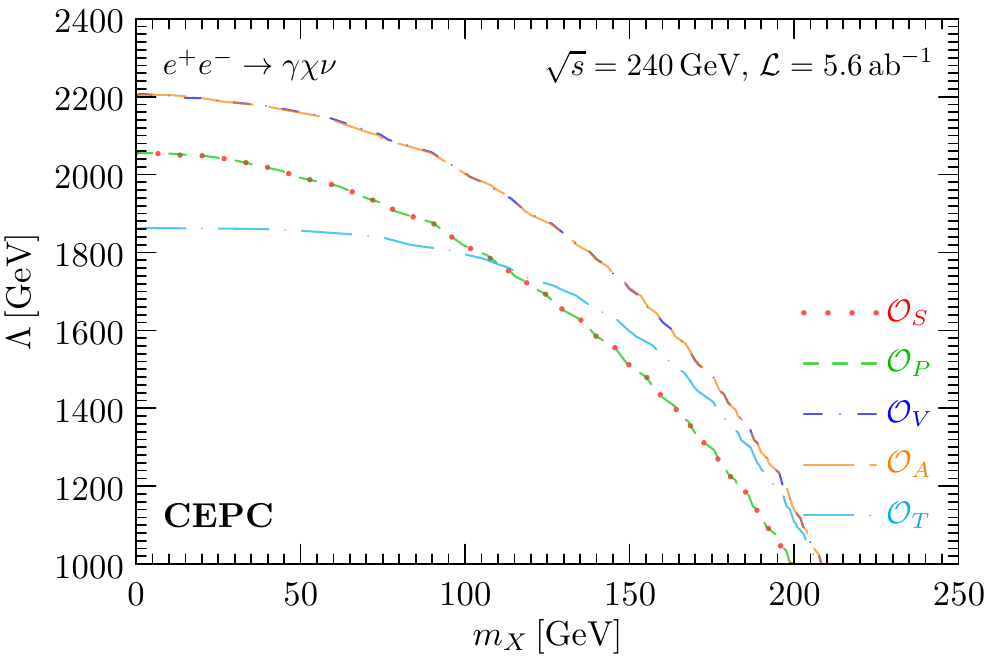}
\includegraphics[width=0.32\textwidth]{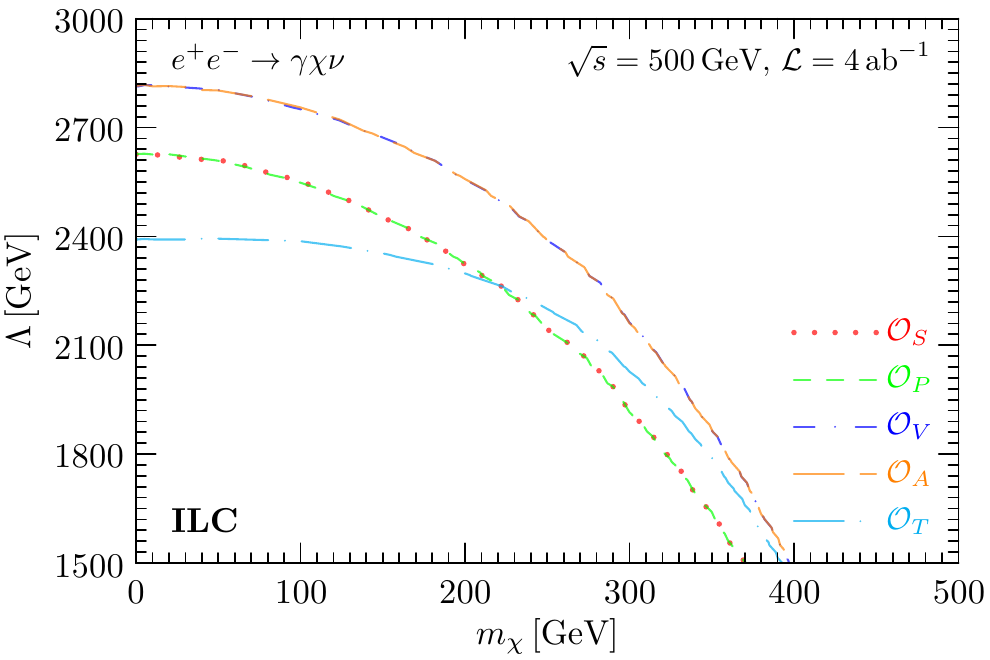}
\includegraphics[width=0.32\textwidth]{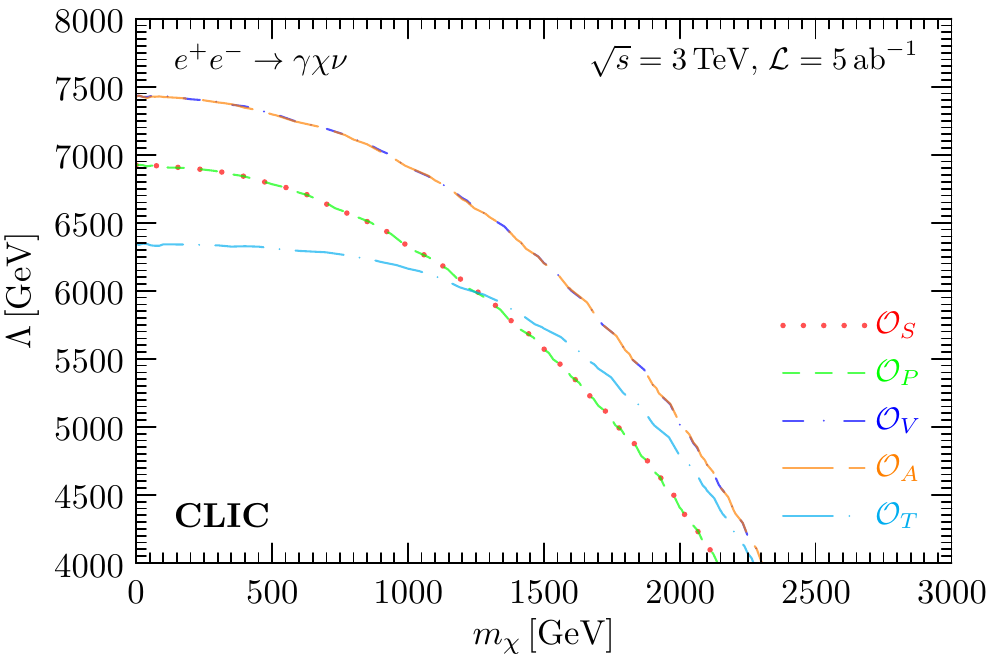}
\caption{\it The 95\% exclusion sensitivities for the mono-photon
search at CEPC (left), ILC (middle), and CLIC (right).}
\label{fig:chi:monoa:unpolbeams}
\end{figure}

\gfig{fig:chi:monoa:unpolbeams} shows the expected 95\%
exclusion limits at CECP (left), ILC (middle), and CLIC
(right) with unpolarized beams.
As we have shown in \gfig{fig:xe-yxs}, total cross sections of the scalar (vector) 
and pseudo-scalar (axial-vector) are the same, and hence the mono-photon
process is unable to distinguish the corresponding Lorentz structures of the effective operators.
We can also see this property in \gfig{fig:chi:monoa:unpolbeams},
where the exclusion curves for scalar (vector) and
pseudo-scalar (axial-vector) completely overlap.
With the vector and axial-vector operators having the largest 
cross sections for a universal cut-off scale $\varLambda$, 
the expected exclusion limits are the strongest ones.
For a massless dark fermion ($m_\chi =0$), the exclusion limits reach
2.2\,TeV, 2.8\,TeV and 7.4\,TeV at $\sqrt{s} = 240\gev$,
500\,GeV, and 3\,TeV, respectively. For comparison, the
exclusion limits for the scalar and pseudo-scalar operators
are slightly lower, reaching 2.0\,TeV, 2.6\,TeV and 6.9\,TeV,
respectively. The tensor operator has the weakest limits
at 1.8\,TeV, 2.4\,TeV, and 6.3\,TeV, respectively.
Due to the phase space suppression, the exclusion
limits reduce
with increasing dark fermion mass. With higher beam energy, the
exclusion limits increase quite significantly.

\begin{figure}[t]
\centering
\includegraphics[width=0.32\textwidth]{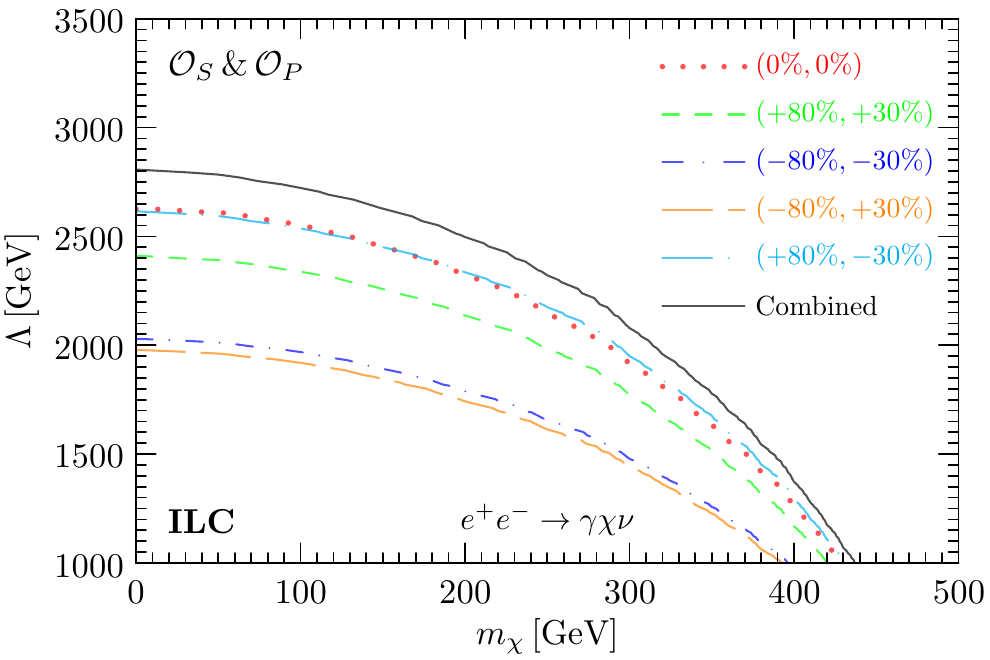}
\hfill%
\includegraphics[width=0.32\textwidth]{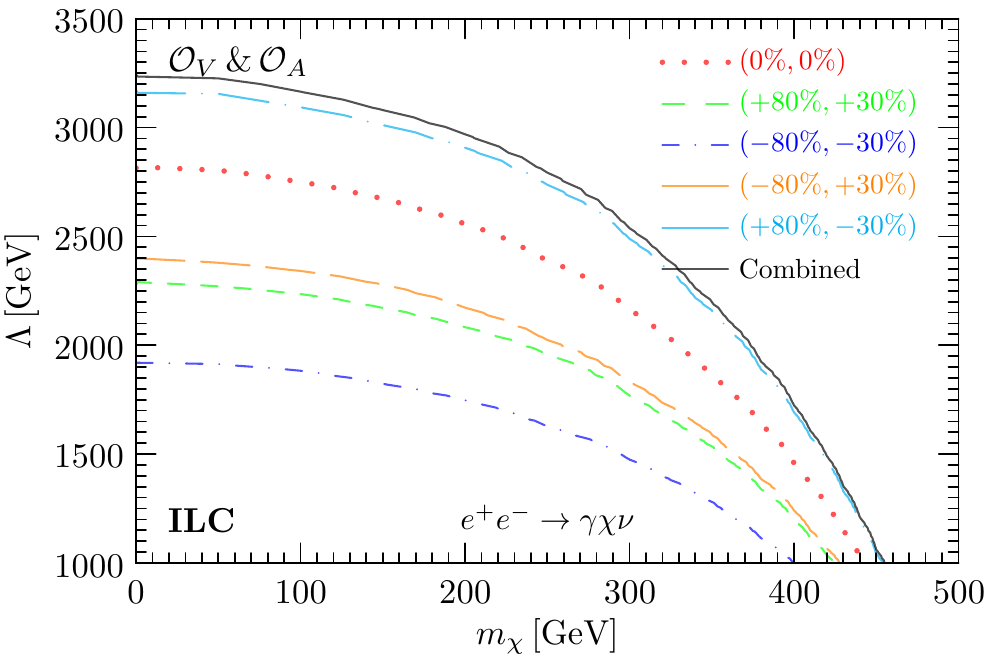}
\hfill%
\includegraphics[width=0.32\textwidth]{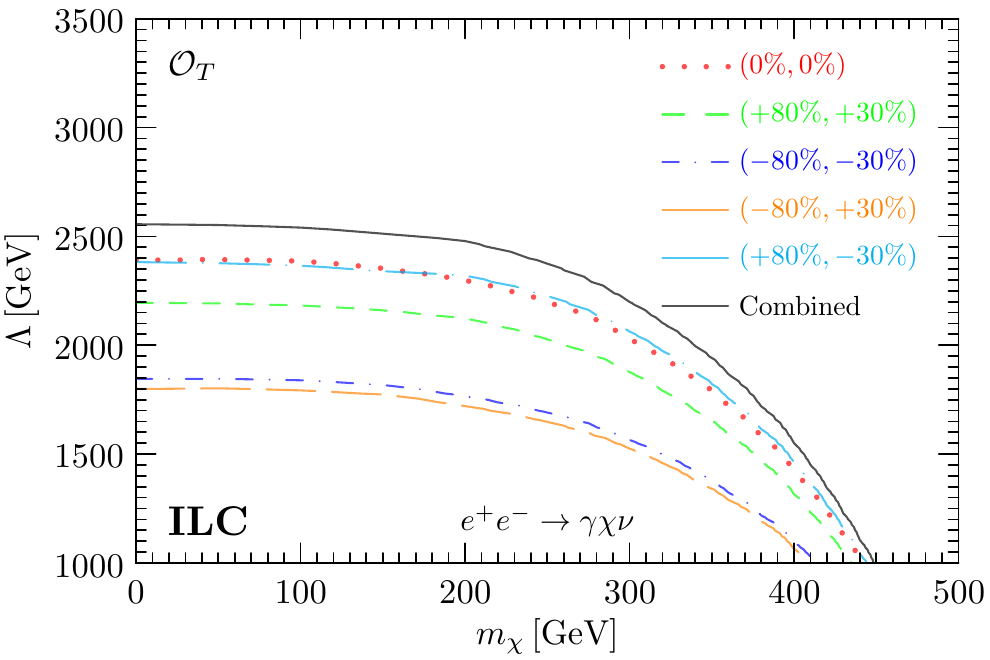}
\caption{\it The expected 95\% exclusion limits provided by the
mono-photon channel at ILC with beam polarizations 
$\big(P_{e^-},\, P_{e^+} \big) = \big(\pm 80\%,\, \pm 30\%\big)$ 
and the corresponding projected luminosities listed in
Table.\,\ref{tab:mechs}. Three cases of (pseudo-)scalar (left), 
(axial-)vector (middle), and tensor (right) operators are
shown. A combined exclusion limit from all the four
polarization configurations are also shown by black-solid line. 
For comparison, the red-dotted lines are for unpolarized beams
while the black solid lines for combined 
polarization configurations.}
\label{fig:chi:monoa:polbeams}
\end{figure}

The effect of beam polarizations are studied at $\sqrt{s}=500\gev$
with typical beam polarizations $P_{e^-} = \pm 80\%$ and 
$P_{e^+} = \pm 30\%$ at ILC for illustration. Using the luminosities
listed in \gtab{tab:mechs}, \gfig{fig:chi:monoa:polbeams} shows
the expected 95\% exclusion limits for the (pseudo-)scalar (left),
(axial-)vector (middle), and tensor (right) operators. 
For comparison, the exclusion limits with unpolarized beams
are also shown by red-dotted lines.
Of the four different beam polarization configurations,
$\big( P_{e^-}, P_{e^+}\big) = (+80\%,\, -30\%)$ is the most
sensitive one as expected. This is particularly useful for the
(axial-)vector operator. In addition, the combined exclusion
limit  (black-solid) from all the four polarization configurations
provides further enhancement.

\section{Electron-Positron Pair Production Associated with Missing Energy}
\label{sec:epemx}

\begin{figure}[t]
\centering
\includegraphics[width=0.235\textwidth]{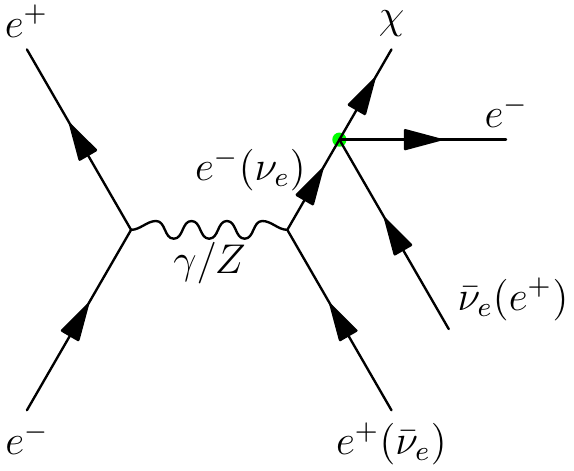}
\hfill%
\includegraphics[width=0.235\textwidth]{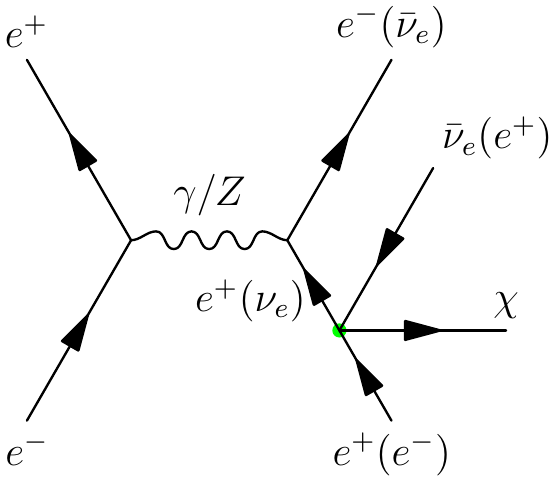}
\hfill%
\includegraphics[width=0.163\textwidth]{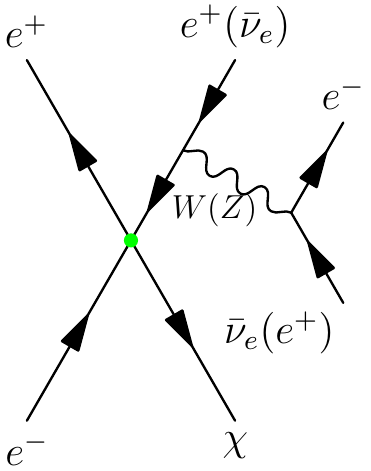}
\hfill%
\includegraphics[width=0.12\textwidth]{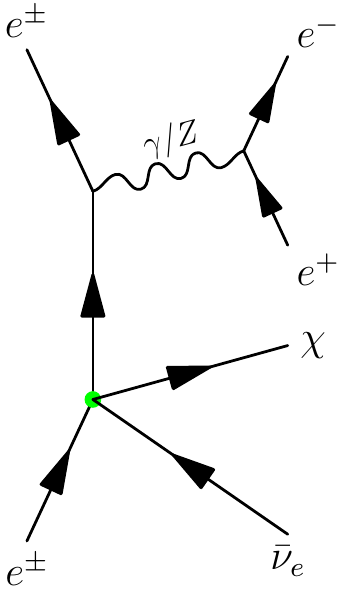}
\hfill%
\includegraphics[width=0.12\textwidth]{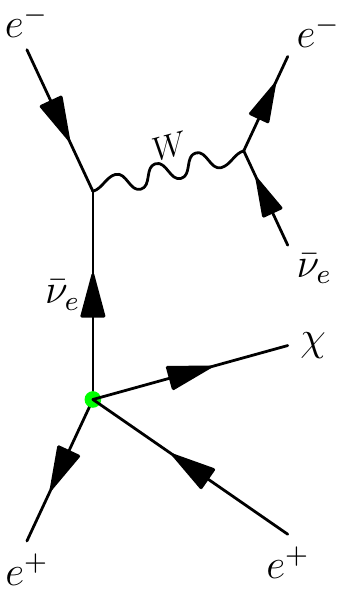}
\\[3mm]
\includegraphics[width=0.18\textwidth]{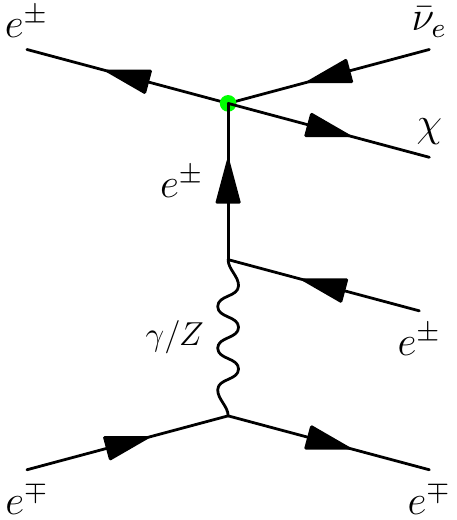}
\hfill%
\includegraphics[width=0.18\textwidth]{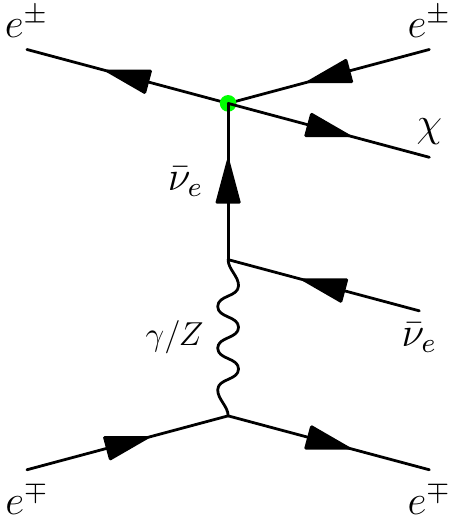}
\hfill%
\includegraphics[width=0.18\textwidth]{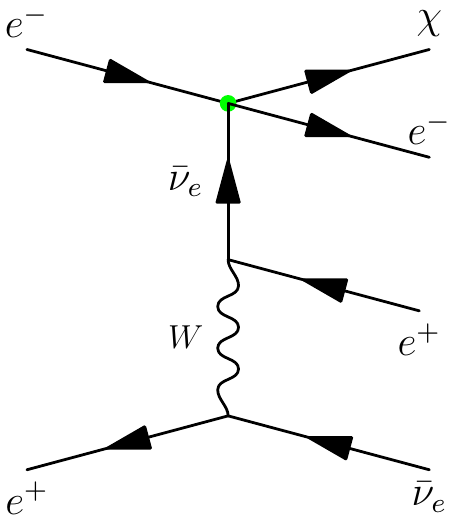}
\hfill%
\includegraphics[width=0.18\textwidth]{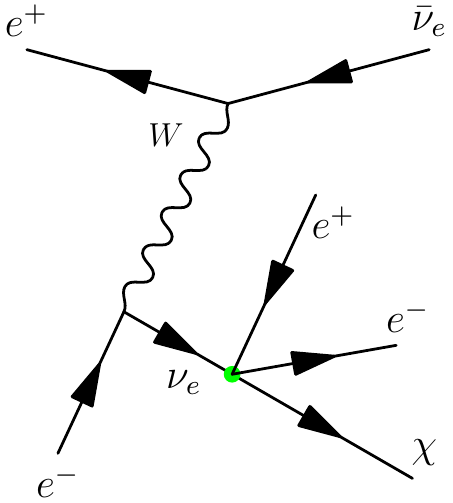}
\hfill%
\includegraphics[width=0.18\textwidth]{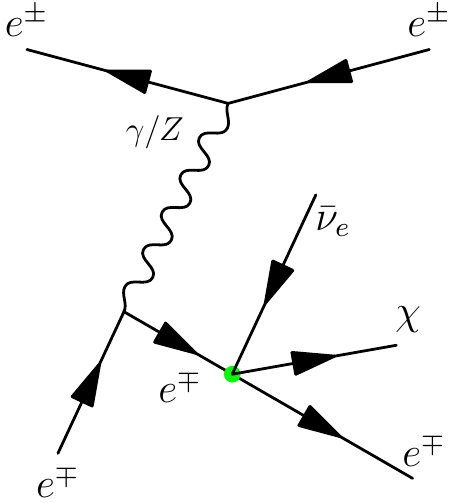}
\caption{\it
The signal Feynman diagrams for the electron-positron
pair production associated with missing energy,
$e^+ e^- \rightarrow e^+ e^- \met$.
}
\label{fig:epm:Feyn}
\end{figure}

\begin{figure}[b]
\centering
\includegraphics[width=0.282\textwidth]{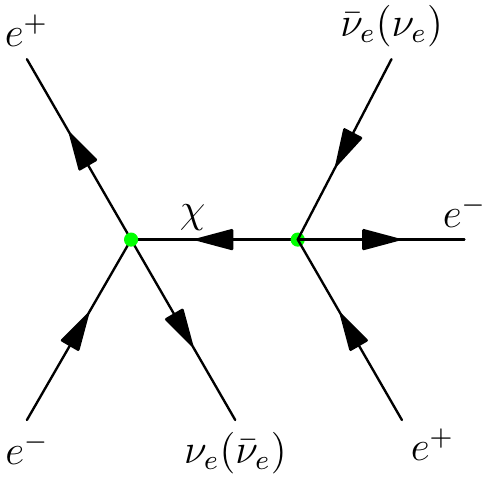}
\hspace{20mm}
\includegraphics[width=0.18\textwidth]{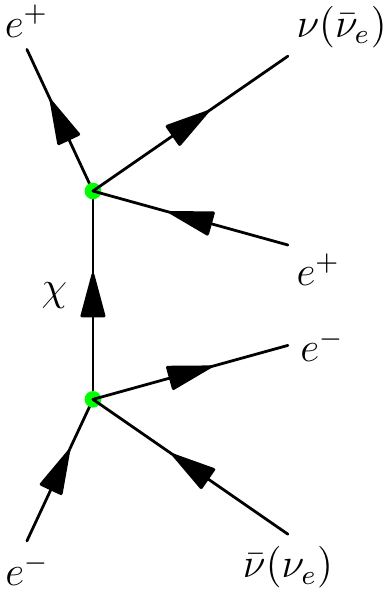}
\caption{\it The signal Feynman diagrams for
the electron-pair production with missing energy carried
away by a neutrino pair, $e^+ e^- \rightarrow e^+ e^- \nu \bar \nu$,
that can interfere with the SM backgrounds.}
\label{fig:epm:interf:Feyn}
\end{figure}

In addition to photon, the charged electron/positron that appears
in the absorption operators is also
detectable at the future $e^+ e^-$ colliders. For example,
the electron-positron pair production associated with
missing energy $e^+ e^- \rightarrow e^+ e^- + \met$ can
also probe the $\chi \nu e^+ e^-$ coupling. Since both the
dark fermion $\chi$ and neutrino are invisible,
two channels can contribute,
\begin{subequations}
\begin{align}
  e^{+} + e^{-}
& \to
  e^{+} + e^{-} + \nu_e + \bar{\chi}\, (\chi + \bar \nu_e)\,,
\\
  e^{+} + e^{-}
& \to e^{+} + e^{-} + \nu_e + \bar \nu_e,
\end{align}
\label{eq:eeETmis}
\end{subequations}
whose
Feynman diagrams are shown in \gfig{fig:epm:Feyn} and
\gfig{fig:epm:interf:Feyn}, respectively.

For the first channel in \gfig{fig:epm:Feyn},
each diagram contains a single effective operator
labeled by the green dot. So the amplitude contains
a factor $1/\Lambda^2$ and the cross section has
$1/\Lambda^4$. The Feynman diagrams in the first row
each contains an $s$-channel resonance at either
the neutral $\gamma/Z$ or the charged $W$ pole.
For comparison, those in the second row arise
from the $t$-channel processes. All these diagrams
have the dark fermion $\chi$ or $\bar \chi$
in the final state, although only $\chi$ is shown
in \gfig{fig:epm:Feyn} for simplicity.

\begin{figure}[t]
\centering
\includegraphics[width=0.48\textwidth]{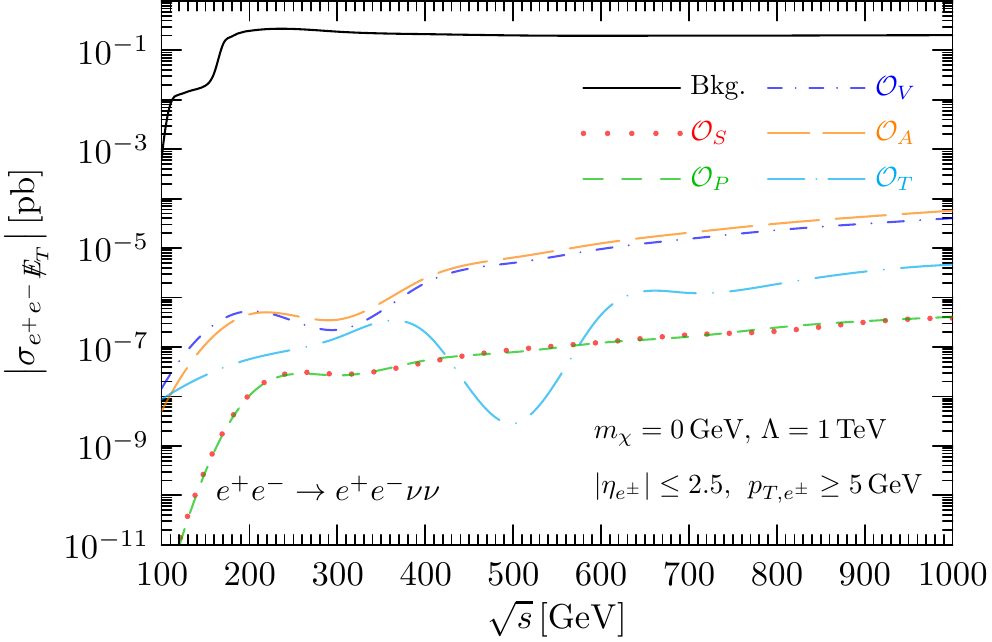}
\hfill%
\includegraphics[width=0.48\textwidth]{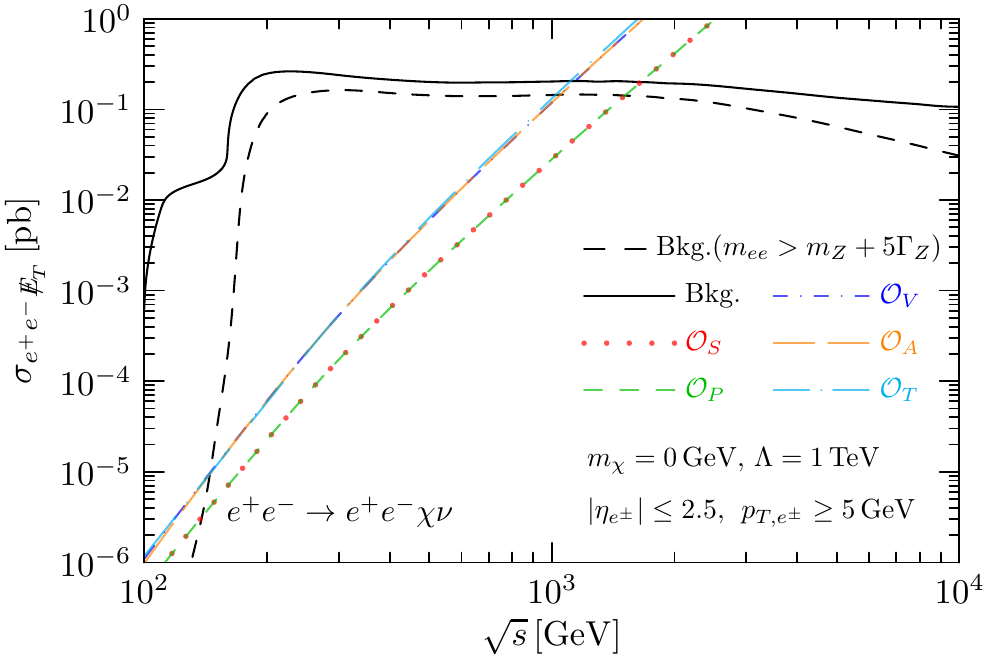}
\caption{\it
{\bf Left panel}: The $e^{+} + e^{-} \to e^{+} + e^{-} + \nu + \bar{\nu}$
cross sections given by purely the SM contribution 
(black-solid) as well as the interference (colorful non-solid) 
between the SM contribution and the effective four-fermion
operators.
{\bf Right panel}: The cross sections of the
$e^{+}e^{-}  + \nu + \bar{\chi}$ and $\chi + \bar{\nu}$
signal process and the purely SM background
$e^{+} e^{-} \to e^{+}e^{-} + 2\nu$.}
\label{fig:XE:YXS:EPX}
\end{figure}

However, missing energy does not mean that $\chi$
has to appear as a final-state particle. Instead, it
can appear as an intermediate particle with neutrino
playing its role to serve as missing energy. To achieve
that, two effective operators are necessary to provide
two $\chi$ lines to form a propagator as shown in
\gfig{fig:epm:interf:Feyn}. Such diagram already receives
a factor of $1/\Lambda^4$ suppression at the amplitude
level. To be comparable with the single-operator
processes in \gfig{fig:epm:Feyn}, an interference with
a SM process that is not suppressed by $\Lambda$ is
required. In total, 56 Feynamn diagrams can contribute
to such SM process with $e^+ e^- \nu_e \bar \nu_e$ final
states whose Feynman diagrams are omitted for simplicity
but can be easily generated with various tools on market
such as \texttt{CompHEP} \cite{CompHEP:2004qpa}, 
\texttt{CalcHEP} \cite{Belyaev:2012qa}, 
and \texttt{MadGraph5} \cite{Alwall:2014hca}. However,
since the SM processes are dominated by nearly
on-shell intermediate vector bosons ($Z$ and $W^\pm$),
the interference effects are significantly reduced
with limited phase space. The SM background is shown
as black lines in \gfig{fig:XE:YXS:EPX}. Note that
the two sudden increases around
$\sqrt s \approx (100 \sim 200) \gev$ come from
the $W$ and $Z$ pair production thresholds.

The left panel of \gfig{fig:XE:YXS:EPX} shows the
interference contributions (colofull non-solid curves) 
in comparison with the corresponding pure SM background
(black-solid). Note that the sign of the interference
contribution depends on the collision energy $\sqrt s$
and the Lorentz structure of the effective four-fermion
operators. For simplicity, only the magnitudes of the
interference contributions are shown without their signs.
One can clearly see that even though the signal
cross sections (with $\varLambda=1\tev$ and $m_\chi = 0\gev$) 
grow with $\sqrt{s}$, it is still more than 3 orders of magnitude
smaller than the background at $\sqrt{s}=1\tev$.

For comparison, the cross sections of the purely
new physics channels
in \gfig{fig:epm:Feyn} grow rapidly with $\sqrt{s}$,
as shown in the right panel of \gfig{fig:XE:YXS:EPX}.
At $\sqrt{s}=1\tev$, their cross sections with
$m_\chi = 0\gev$ and $\varLambda=1\tev$ are about just
1 order of magnitude smaller than and quickly exceed
the SM background for $\sqrt s \gtrsim 1$\,TeV. It is
of great advantages for searching the electron-positron
pair production associated with missing energy at CLIC
that has running modes at 1.5\,TeV and 3\,TeV.
Then, the interference channels in \gfig{fig:epm:interf:Feyn}
can be safely ignored.

Since things are not settled yet to forsee which
future $e^+ e^-$ collider would finally get running,
we also study the projected
sensitivities at CEPC and ILC. Before doing specific
analysis, the $e^+ e^-$ events are selected by some
basic cuts,
\begin{align}
  |\eta_{e^\pm}| < 2.5,
\qquad
  p_{T, e^\pm} > 5\gev,
\label{eq:eeEtCuts}
\end{align}
which can universally apply for CEPC, FCC-ee, ILC,
and CLIC. Below we will show more details about this
process.

\subsection{Differential Cross Sections for Distinguishing Signals and Background}

\begin{figure}[t]
\centering
\includegraphics[width=0.32\textwidth]{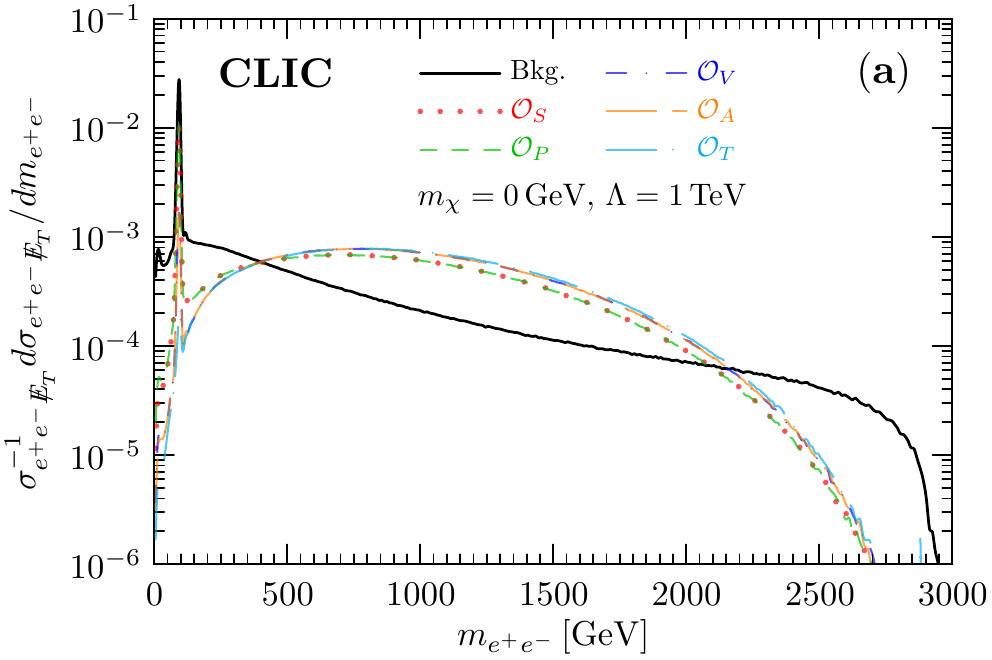}
\includegraphics[width=0.32\textwidth]{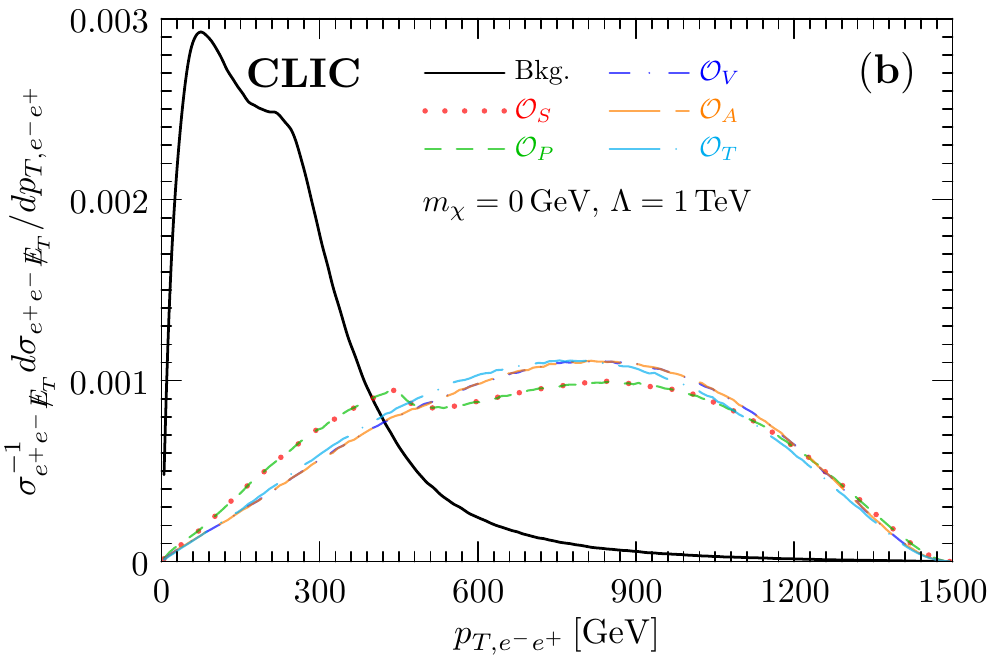}
\includegraphics[width=0.32\textwidth]{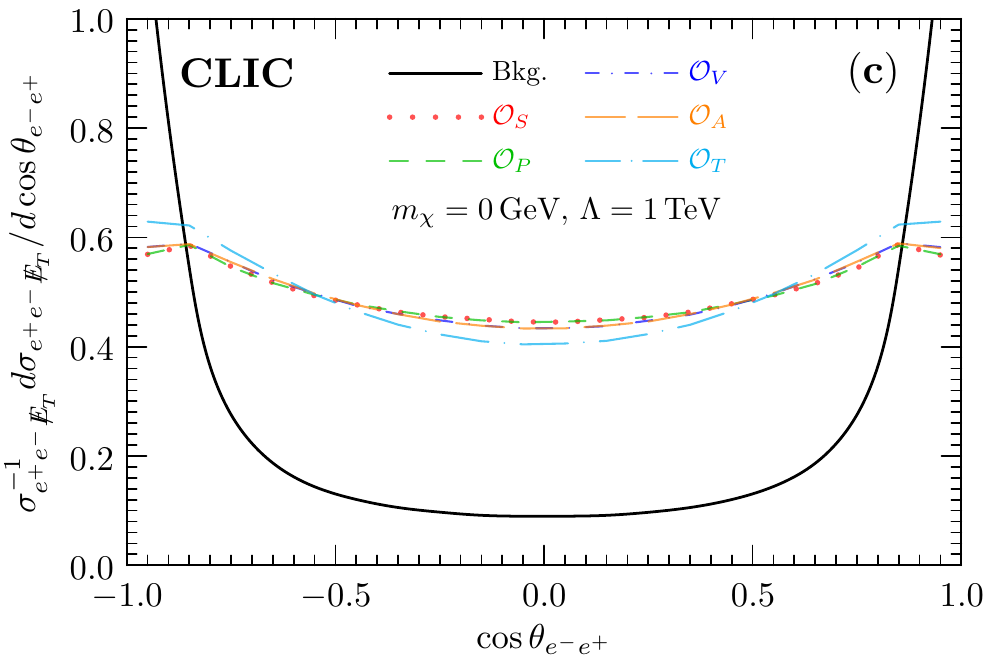}
\\[2mm]
\includegraphics[width=0.32\textwidth]{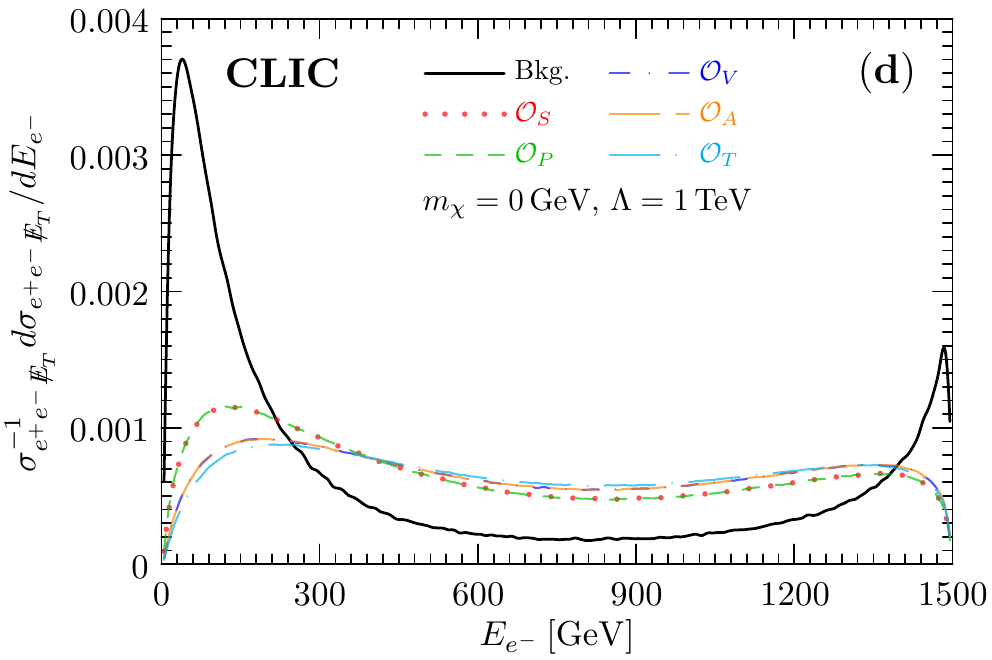}
\includegraphics[width=0.32\textwidth]{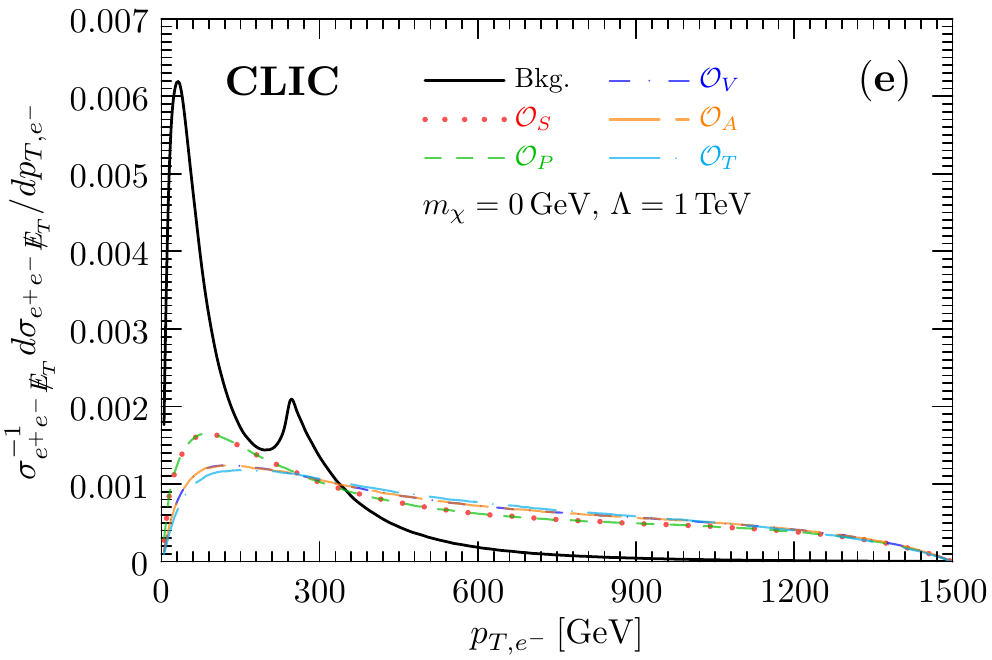}
\includegraphics[width=0.32\textwidth]{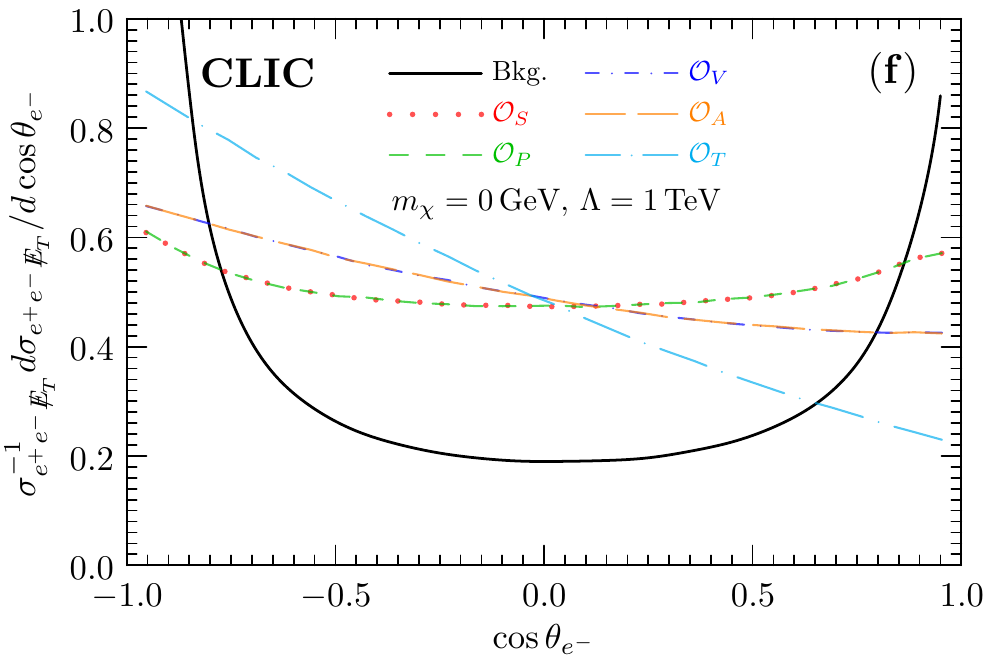}
\\[2mm]
\includegraphics[width=0.32\textwidth]{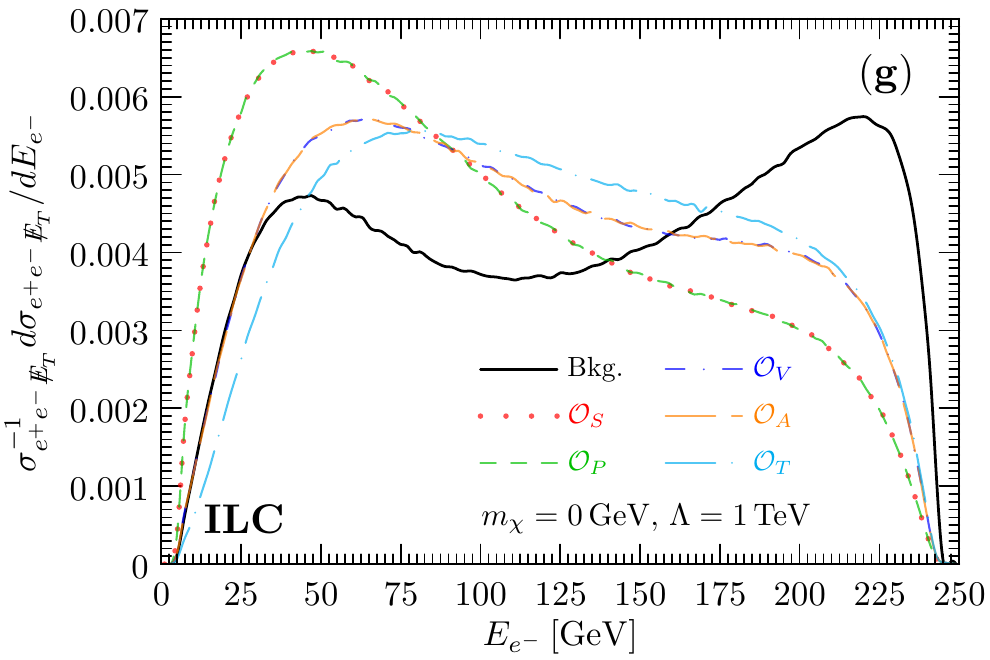}
\includegraphics[width=0.32\textwidth]{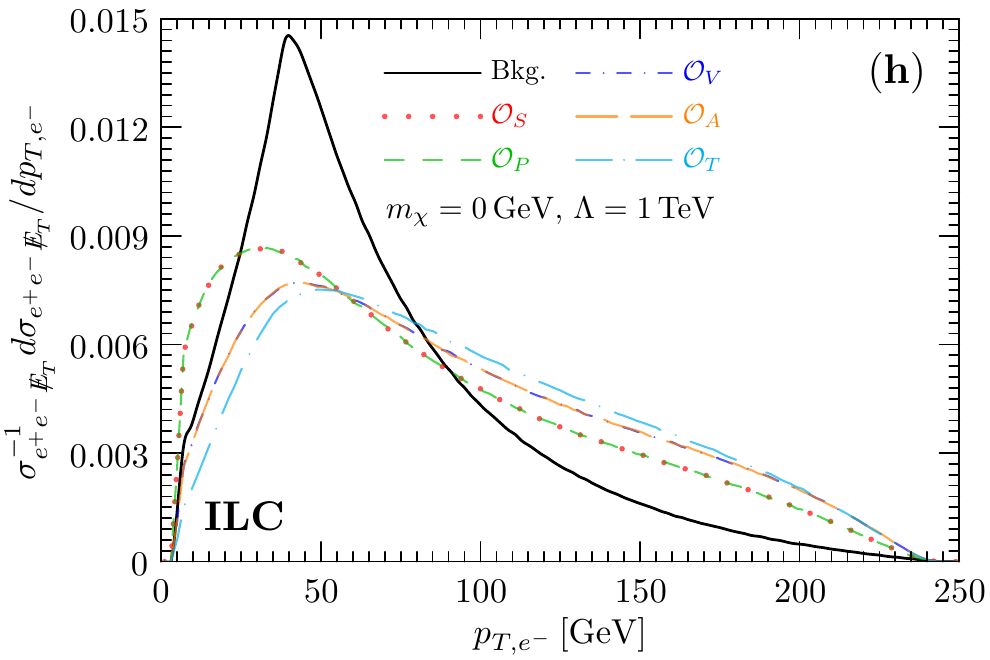}
\includegraphics[width=0.32\textwidth]{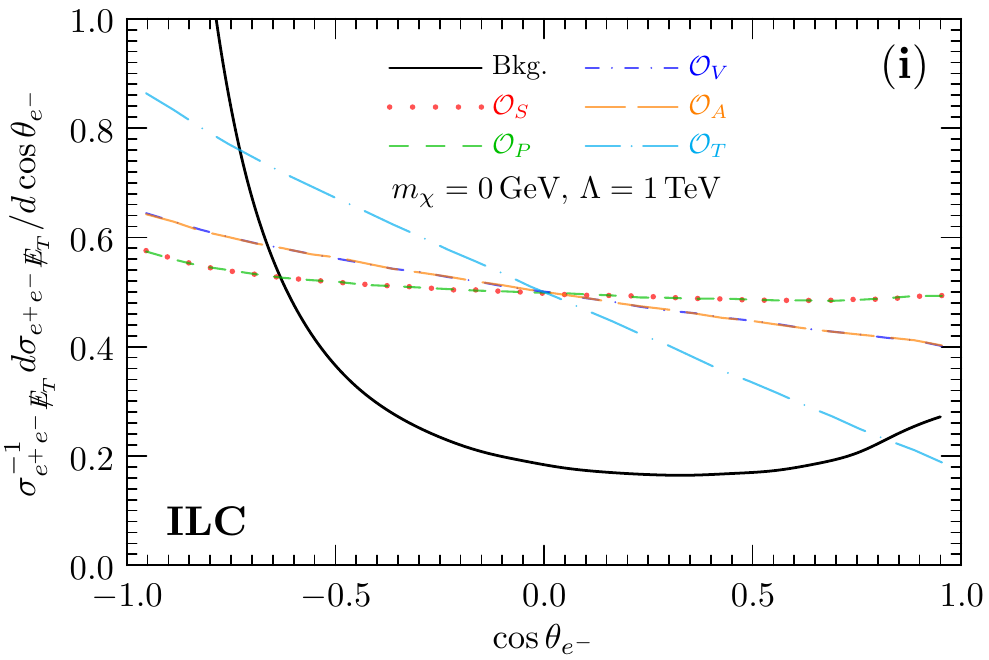}
\caption{\it The differential cross sections of the
$e^{+} e^{-} \to e^{+} e^{-} + \met$ channel.
For signals (colorful non-solid curves), only 
$e^{+}e^{-}  + \nu + \bar{\chi}\, (\chi + \bar{\nu})$
are taken into account while the black-solid curves
show the SM background $e^{+} e^{-} \to e^{+} e^{-} + 2\nu$.
}
\label{fig:epx:kin}
\end{figure}

\gfig{fig:epx:kin} shows some differential distributions
of the $e^+ e^- \rightarrow e^+ e^- \met$ process at
CLIC with $\sqrt{s} = 3\tev$ and selected by
kinematic cuts in \geqn{eq:eeEtCuts}.
Both signals and background have a $Z$ resonance peak
in the invariant mass $m_{e^+e^-}$ spectrum, which are
shown as \gfig{fig:epx:kin}(a). While the
signal has a bump in the range of
$500\gev \lesssim m_{e^+ e^-} \lesssim 2000\gev$,
the SM background curve is more flat. In other words,
the middle region of the $m_{e^+ e^-}$ distribution
provides more signal significance than the region
around the $Z$ resonance.

However, the $Z$ resonance in the signal and background
events are quite different. For background, the $Z$
bosons are pairly producted via $t$-channel ISR process
and then decay into electron-positron and neutrino pairs,
$e^+e^- \to Z + Z \to 2e + 2\nu$. Such on-shell
$Z$ bosons move in the forward and backward regions,
as can be seen in \gfig{fig:epx:kin}(c). In addition,
the transverse momentum $p_{T, e^+e^-}$ of the
$e^+ e^-$ system tends to have small values as
shown in \gfig{fig:epx:kin}(b). In contrast, the
signal $Z$ resonance is not that forwarded/backwarded
and hence the electron-positron pair can have larger
transverse momentum. These features can be used to
distinguish signals from the SM background.

Among various operators, the differential distributions
are not that different. The kinematic distributions
of the $e^+ e^-$ system as a whole cannot help much
to distinguish different signals. This is because
the electron bilinear in the effective operators
\geqn{eq:operators} behaves like an effective
particle. For example, $\bar e e$
($\bar e \gamma_5 e$) and $\bar e \gamma_\mu e$
($\bar e \gamma_\mu \gamma_5 e$) for the
$e^+ e^-$ system can be treated as a scalar
and vector particle, respectively.
The kinematic observable such as
$m_{e^+ e^-}$, $p_{T, e^+ e^-}$, or 
$\cos \theta_{e^+ e^-}$ apply for both
scalar and vector particles. Without reconstructing
the total spin of the $e^+ e^-$ system, there
is no way to distinguish different operators.
The same logic also works for the tensor operator.

Fortunately, the situation can be improved by
using the distributions of the individual
final-state electron or positron that are shown
in the second and third rows of \gfig{fig:epx:kin}. 
Comparing the situations at CLIC (middle row)
and ILC (buttom row), the difference among signals
is larger with smaller collision energy.
CLIC has large enough difference in only the
$\cos \theta_{e^-}$ distribution. In contrast,
the $E_{e^-}$ and $p_{T, e^-}$ distributions
are also quite different at ILC. It is of great
advantages for the $e^+ e^- \met$ search at future
lepton colliders to overcome the shortcoming of
the mono-photon channel.

For the electron $p_T$ distributions, the
background has much narrower peak than signals.
This is because both the background $e^+ e^-$ and
$\nu \bar \nu$ systems arise from the $Z$
resonance. Consequently, the electron pair
invariant mass $m^2_{e^+ e^-} = m^2_Z$ is
fixed by the $Z$ boson mass. In the rest frame
of $Z$, electron carries energy $m_Z/2$ and its
angular spectrum follows $d \cos \theta$ with
even distribution for $\cos \theta$. Since
$p_T \propto \sin \theta$, its probability
distribution scales as $\tan \theta$ which
diverges at $\theta \rightarrow 90^\circ$. In other
words, the electron transverse momentum peaks
at $m_Z/2$ in the $Z$ rest frame. Since the
ISR $Z$ bosons move in the forward and
backward regions, the electron transverse
momentum has roughly the same distribution
in the lab frame. That explains why the black
curve in \gfig{fig:epx:kin}(h) has a peak
around $m_Z/2$. The same logic applies for
the neutrino pair system,
$m^2_{\nu \bar \nu} = m^2_Z$, which carries
an energy of at least $m_Z$. For comparison,
the signal $\chi \nu$ system can have larger
energy to allow the colorful lines in
\gfig{fig:epx:kin}(h) to have more events
at the higher end of the $p_{T, e^-}$
distribution.

With more spectrum features than the mono-photon
search, especially the apparent difference between
signals and background, it makes much more sense to
employ binned $\chi^2$ analysis. We choose the
invariant mass of the electron pair ($m_{e^+e^-}$),
the electron energy ($E_{e^-}$) and transverse momentum 
($p_{T, e^-}$) as three representative observables.
The experimental significance is estimated by the
following $\chi^2$,
\bee
  \chi^2 
= 
  \sum_{a}\sum_{i}
\frac{ \big( N^{\rm Sig}_i ( \calo_a ) \big)^2 }
{ N^{\rm Bkg}_i( \calo_a ) + N^{\rm Sig}_i( \calo_a ) }.
\ene
Here $\calo_a = m_{e^+e^-}$, $E_{e^-}$, and $p_{T, e^-}$
are the representative observables while
$N^{\rm Bkg}_i( \calo_a )$ and $N^{\rm Sig}_i( \calo_a )$
are the background and signal event numbers
in the $i$-th bin of the observable $\calo_a$. 

Three typical future $e^+ e^-$ colliders
(CEPC, ILC, and CLIC) are considered.
For CEPC, the detector can identify the prompt 
leptons with high efficiency and purity
\cite{CEPCStudyGroup:2018ghi}. 
For lepton with energies above 5\,GeV, the
identification efficiency can be higher than 99\% 
and the misidentification rate smaller than 2\%.
Similar thing happens at ILC for the tracking
efficiency that can reach  $\sim100\%$ for
a charged particle with momentum $p > 1\gev$
\cite{Bambade:2019fyw}. Therefore, it is
safe to assume 100\% efficiency for 
the electron and positron reconstructions
in the following analysis.

\begin{figure}[t]
\centering
\includegraphics[width=0.32\textwidth]{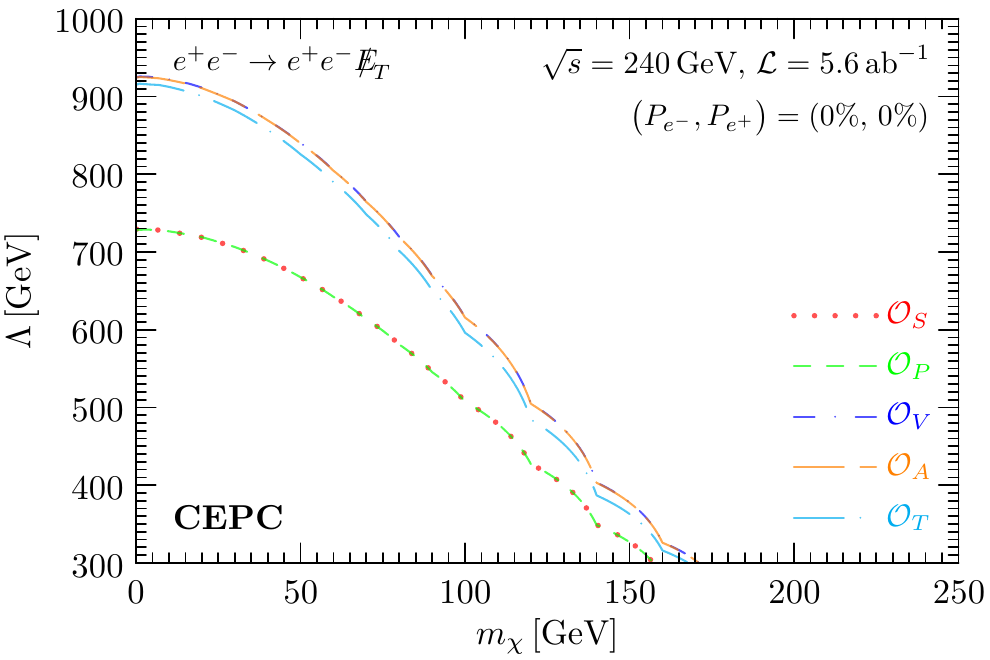}
\includegraphics[width=0.32\textwidth]{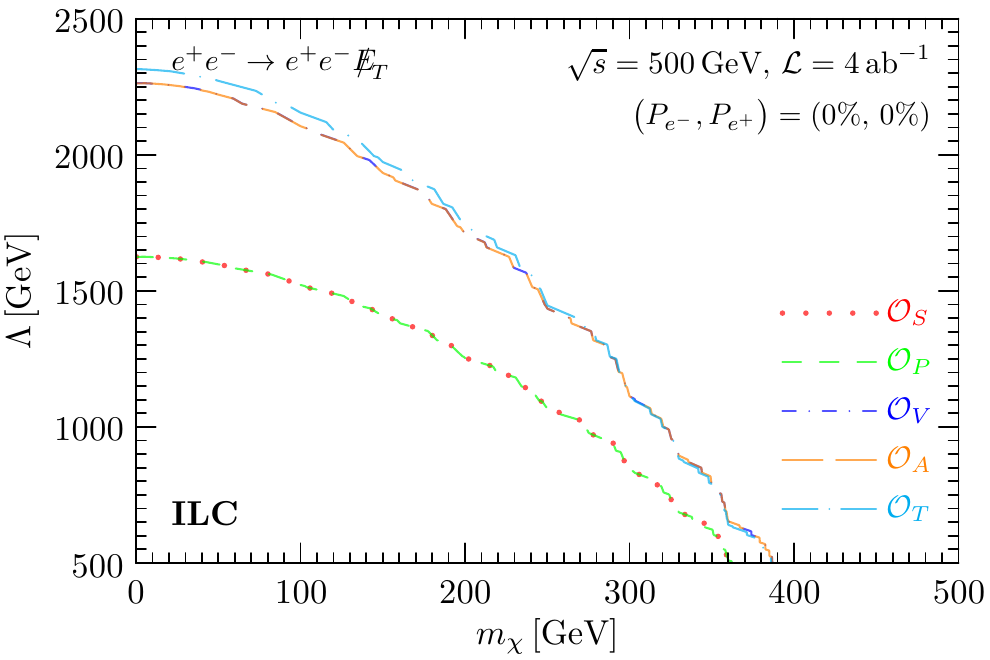}
\includegraphics[width=0.32\textwidth]{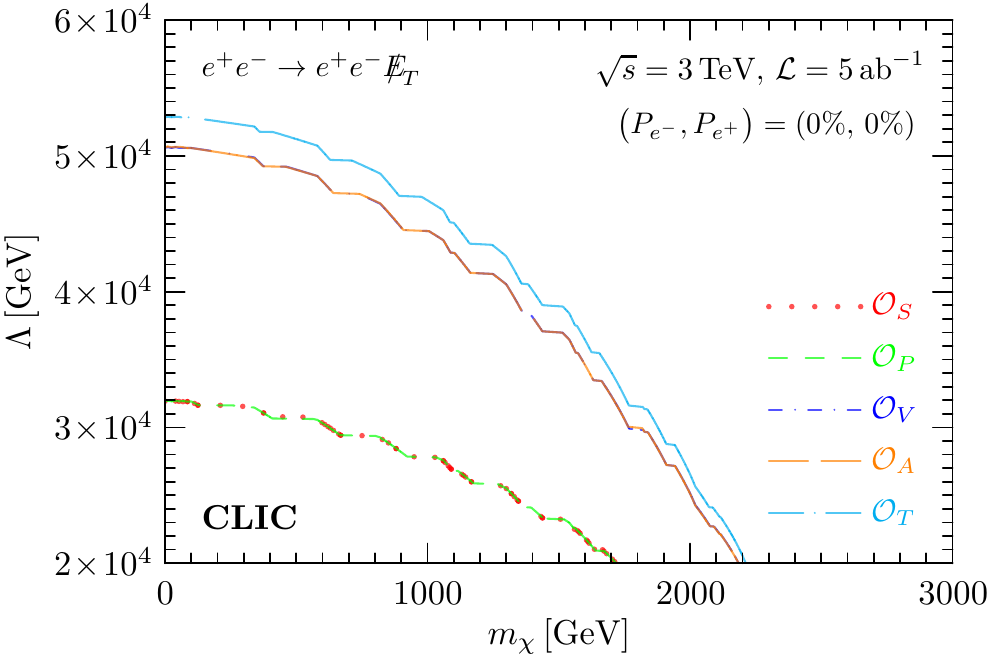}
\caption{\it The expected 95\% exclusion limits by
the $e^+e^-\to e^+e^- + \met$ channel at CEPC (left),
ILC (middle), and CLIC (right).}
\label{fig:chi:epx:unpolbeams}
\end{figure}
\gfig{fig:chi:epx:unpolbeams} shows the expected
95\% exclusion limits at CECP (left), ILC (middle),
and CLIC (right) with unpolarized beams.
For all three colliders, the (axial-)vector and
tensor operators have the strongest exclusion
limits while the (pseudo-)scalar ones are 
weaker. As expected, the limits increase with
the collision energy. With the limits at CEPC
being below 1\,TeV, ILC can reach about 2.2\,TeV,
1.6\,TeV, and 2.3\,TeV for the (axial-)vector,
(pseudo-)scalar, and tensor operators,
respectively. The limits further enhance to
about 51\,TeV, 32\,TeV, and 53\,TeV at CLIC.
Comparing with the mono-photon search in
\gsec{sec:monoa}, the $e^+ e^- \met$ channel
has lower sensitivity at CEPC but can double
the values at CLIC due to the fast increase
of cross section with the collision energy $\sqrt s$.

\subsection{Beam Polarizations for Further Suppressing Background}

\begin{figure}[t]
\centering
\includegraphics[width=0.48\textwidth]{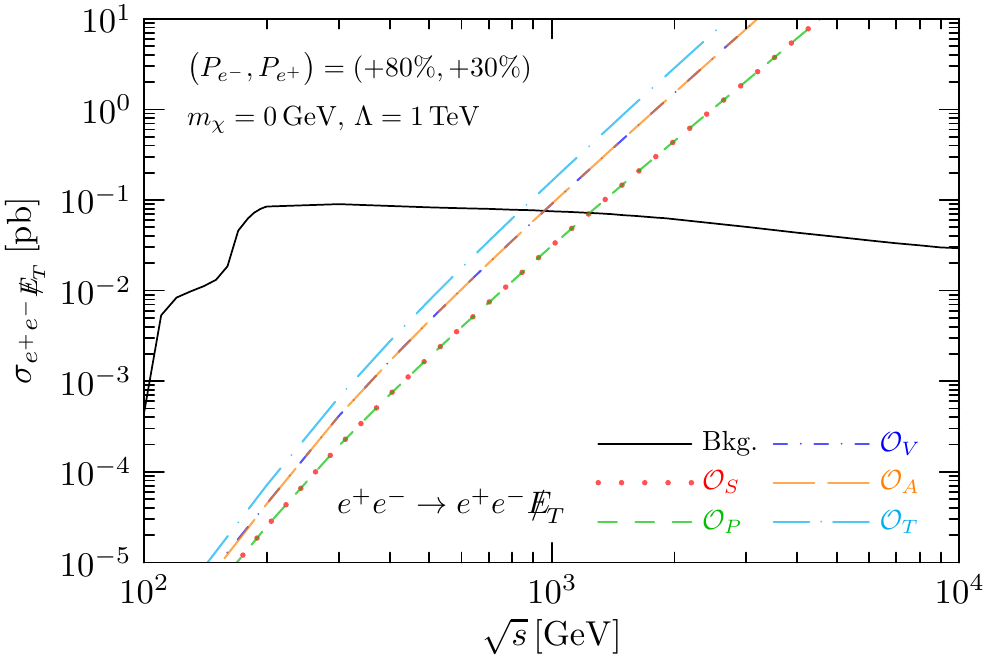}
\quad
\includegraphics[width=0.48\textwidth]{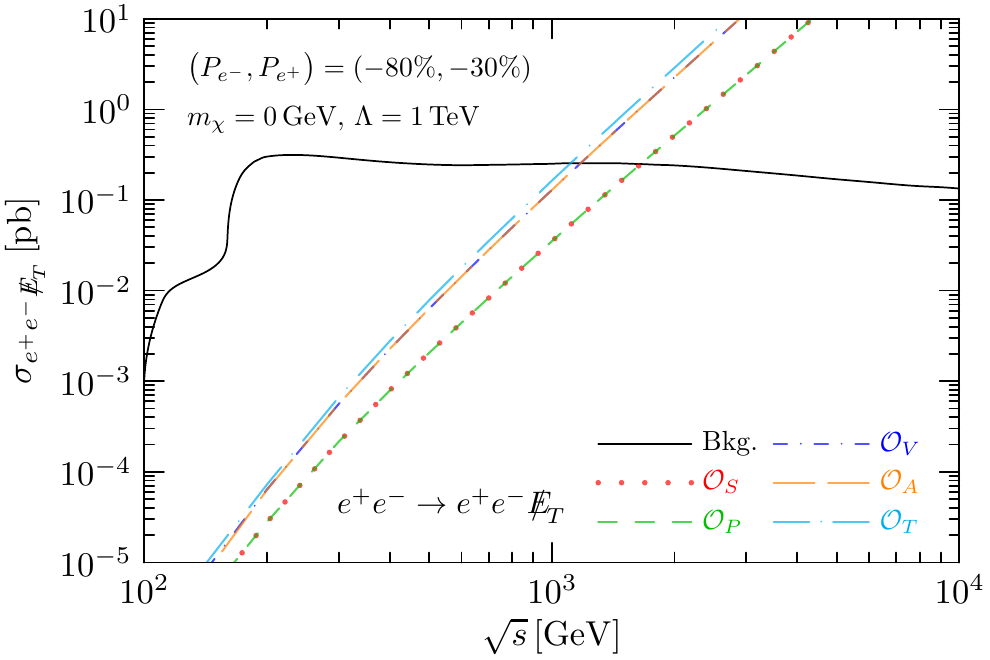}
\\[2mm]
\includegraphics[width=0.48\textwidth]{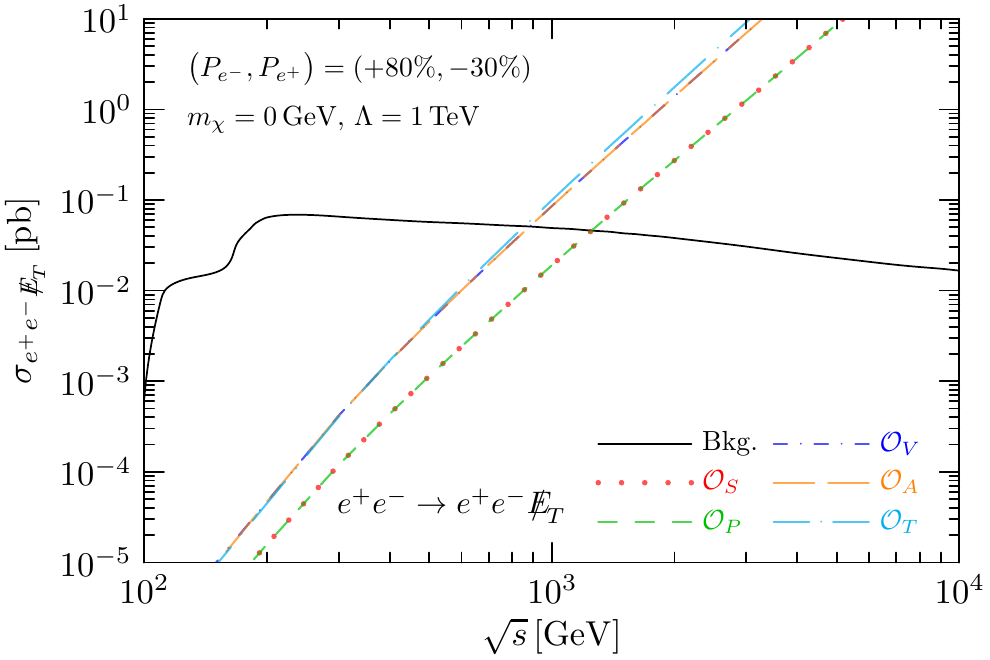}
\quad
\includegraphics[width=0.48\textwidth]{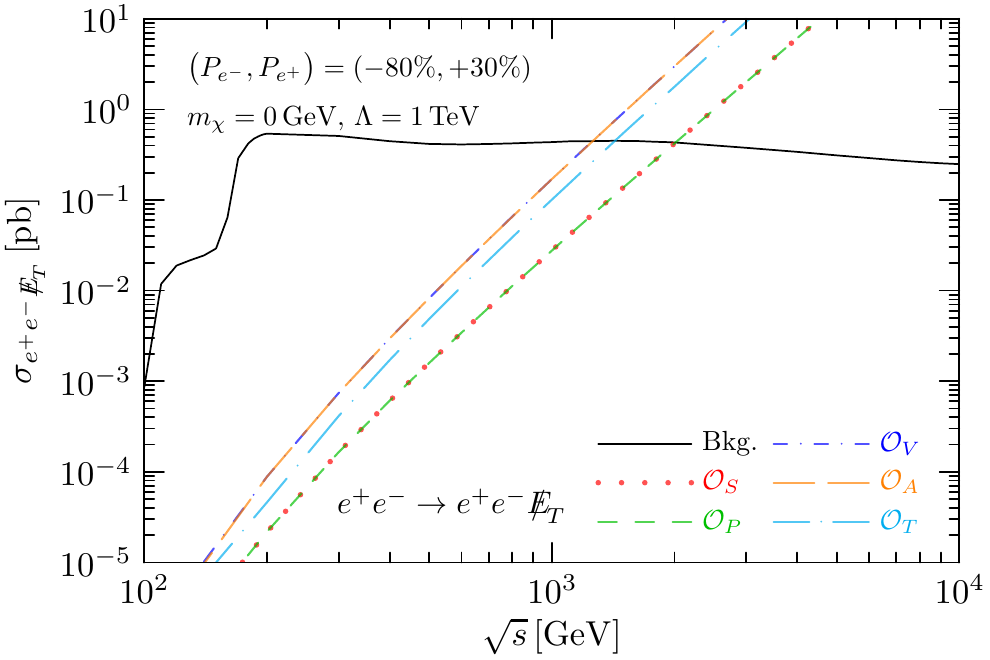}
\caption{\it The
polarized cross sections of the $e^{+}e^{-} \to e^{+}e^{-} + \met$ process
as function of the collision energy $\sqrt{s}$. 
The polarization effects are illustrated with the 
beam polarization configurations $(P_{e^-},\, P_{e^+})=(+80\%, +30\%)$ (top-left panel),
$(-80\%, -30\%)$ (top-right panel), $(+80\%, -30\%)$ (bottom-left panel),
and $(-80\%, +30\%)$ (bottom-right panel), respectively.
While the signals (colorful non-solid curves) are shown
with parameters $m_{\chi}=0\gev$ 
and $\varLambda_i=1\tev$, the black-solid curve shows the 
background from the pure SM contribution to the process 
$e^{+} e^{-} \to e^{+} e^{-} + 2\nu$. All the plots are generated using events passing 
the kinematic cuts $|\eta_{e^\pm}| < 2.5$ and $p_{T, e^\pm} > 5\gev$.
}
\label{fig:xe-yxs:epx:e80p30}
\end{figure}

As discussed in \gsec{sec:beamA}, the beam polarization can be
used to enhance the signal significance using the parity
violation nature of the SM electroweak interactions and
the effective operators in \geqn{eq:operators}.
\gfig{fig:xe-yxs:epx:e80p30} shows the signal and background
cross sections as function of the collision energy $\sqrt s$
for beam polarizations $P_{e^-}=\pm80\%$ and $P_{e^+}=\pm30\%$.
Those backgrounds contributed by the SM left-handed charged
current can be suppressed by right-handed electron beam
as we can see in the top- and bottom-left panels of
\gfig{fig:xe-yxs:epx:e80p30}. Since the SM neutral current
has roughly the same coupling strength with the left- and
right-handed charged leptons, different beam polarizations
make no big difference. Although signals are also reduced,
their decrease is much smaller than background to achieve
better experimental sensitivity. It turns out the
polarization configuration $P_{e^-} = +80\%$ and
$P_{e^+} = -30\%$ gives the lowest background. The
optimal polarization choice here is the same as the
one for the mono-photon search since both have
backgrounds due to the same SM weak interactions.

\begin{figure}[t]
\centering
\includegraphics[width=0.32\textwidth]{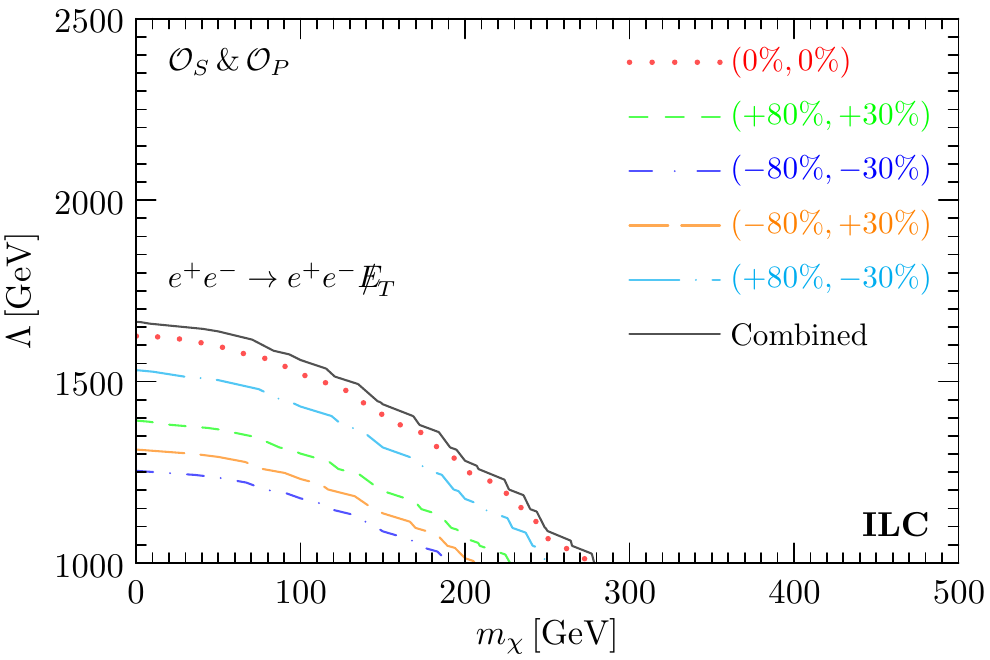}
\includegraphics[width=0.32\textwidth]{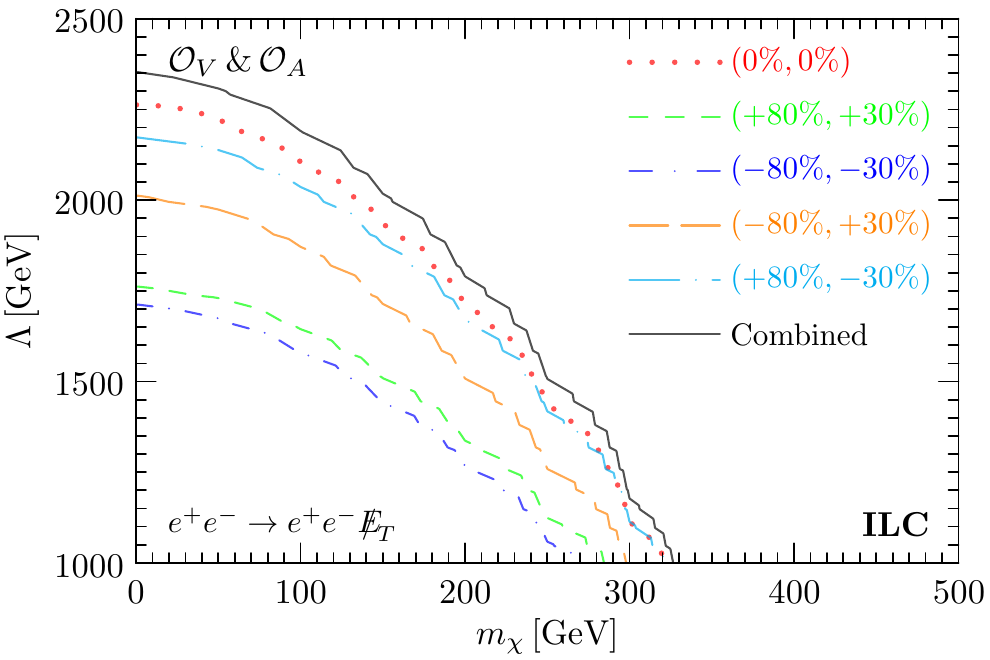}
\includegraphics[width=0.32\textwidth]{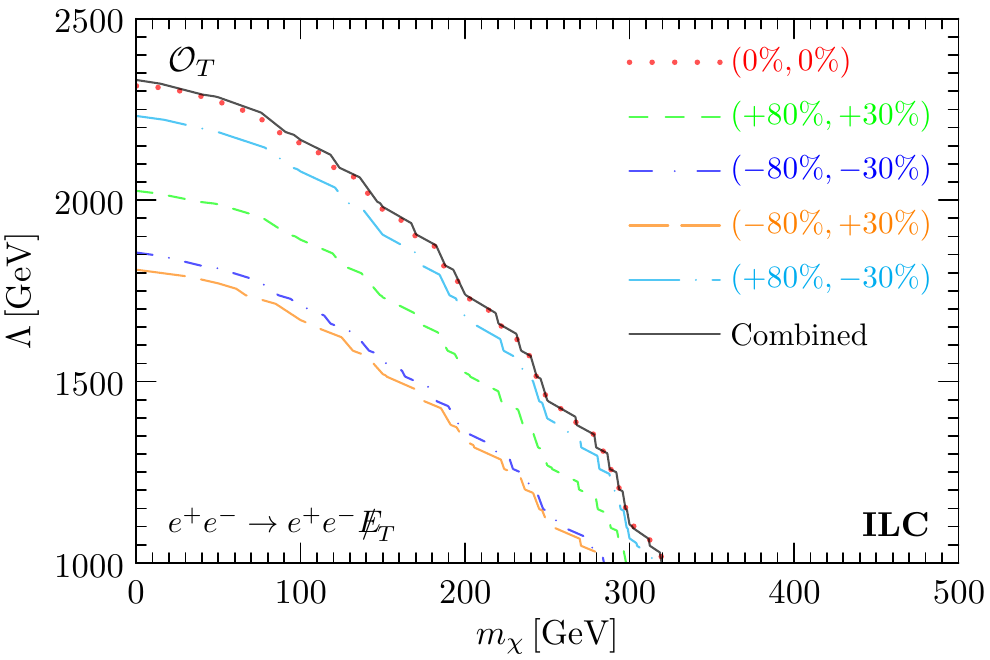}
\caption{\it The
expected exclusion limits at 95\% C.L. by searching for the 
$e^+e^- \to e^+e^- + \met$ process at ILC ($\sqrt{s}=500\gev$)
with beam polarizations 
$\big(P_{e^-},\, P_{e^+} \big) = \big(\pm 80\%,\, \pm 30\%\big)$ 
and the corresponding projected luminosities listed in Table.\,\ref{tab:mechs}. 
The left, middle and right panels show the results for the (pseudo-)scalar, 
(axial-)vector and tensor operators, respectively. A combined exclusion limit 
from all the four polarization configurations is also shown by
the black-solid line. 
For comparison, the exclusion limits with unpolarized beams are also shown 
by red-dotted lines.}
\label{fig:chi:epx:polbeams}
\end{figure}

We illustrate the beam polarization effects for ILC
at $\sqrt{s}=500\gev$ with the typical polarization
configurations $P_{e^-} = \pm 80\%$ and $P_{e^+} = \pm 30\%$.
The expected exclusion limits at 95\% C.L. are shown
in \gfig{fig:chi:epx:polbeams} for (pseudo-)scalar (left),
(axial-)vector (middle), and tensor (right) operators.
With a same luminosity for each polarization configuration,
$\big( P_{e^-}, P_{e^+}\big) = (+80\%,\, -30\%)$ has
the strongest constraint as expected. For comparison, 
the exclusion limits with unpolarized beams are also shown
as red-dotted lines. With smaller luminosity (1.6\,ab$^{-1}$
or 0.4\,ab$^{-1}$ as summarized in \gtab{tab:mechs}),
the exclusion limit for individual configuration is weaker
than the unpolarized one with 4\,ab$^{-1}$ luminosity.
But the combined result with all four configurations
is indeed stronger.

\subsection{Combined Results}

\begin{figure}[t]
\centering
\includegraphics[width=0.32\textwidth]{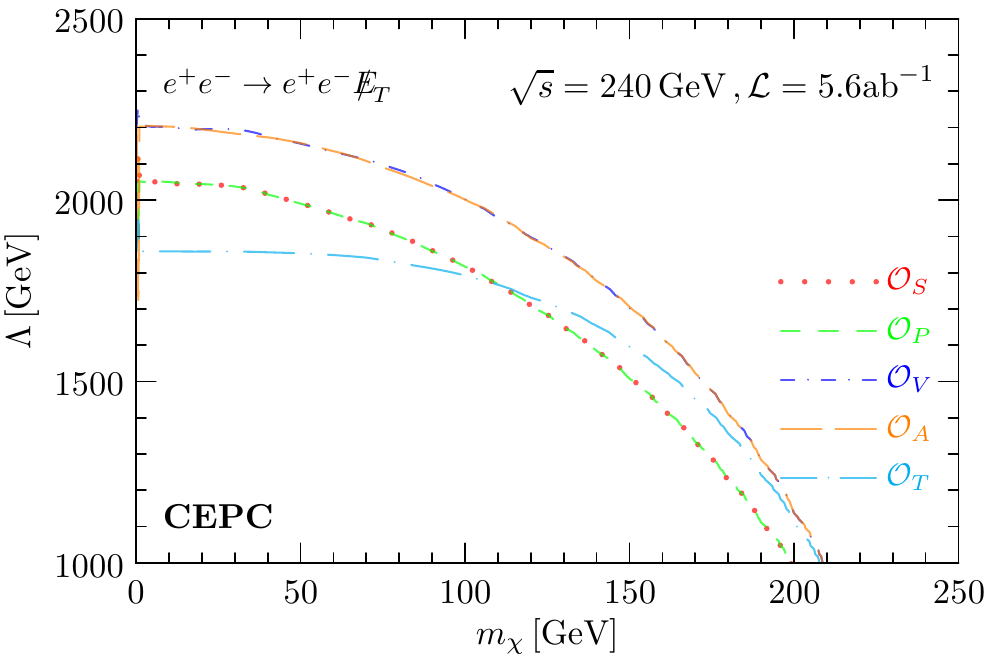}
\hfill%
\includegraphics[width=0.32\textwidth]{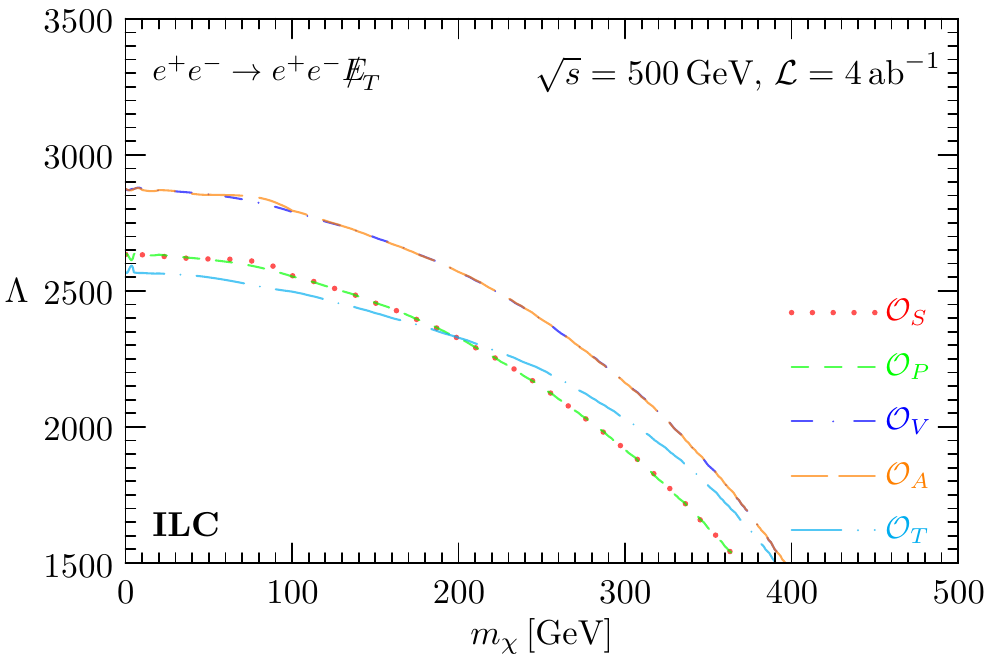}
\hfill%
\includegraphics[width=0.32\textwidth]{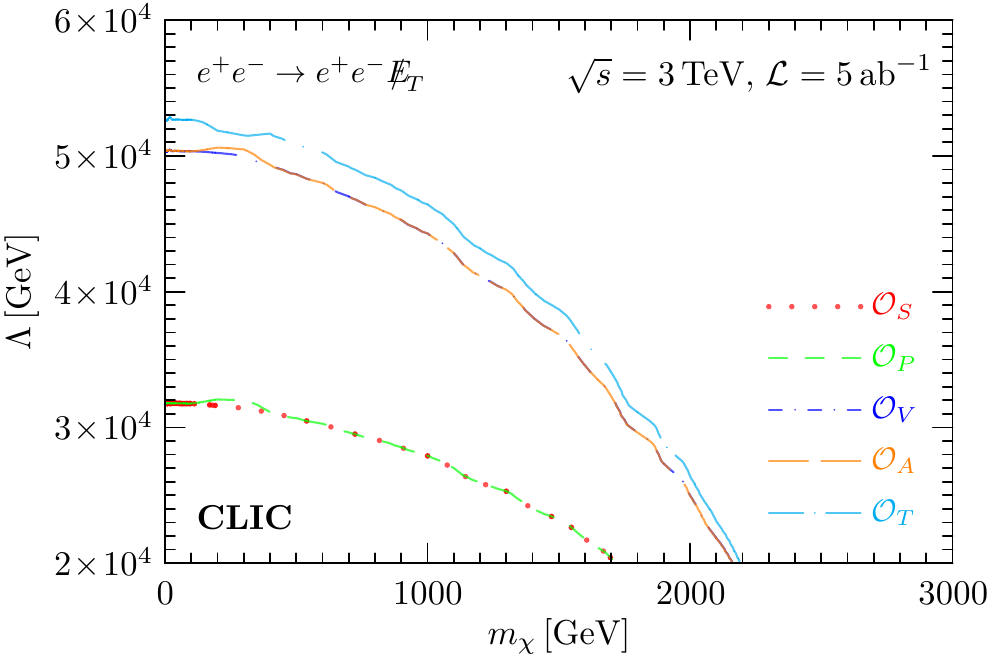}
\caption{\it 
The combined exclusion limits at 95\% C.L. with both
the mono-photon and the $e^{+}e^{-} \to e^{+}e^{-} + \met$
searches at CEPC (left), ILC (middle), and CLIC (right).
For each detector, we have implemented their projected
luminosities as summarized in Table.\,\ref{tab:mechs}.
As a conservative estimation, no beam polarization is
considered.}
\label{fig:chi:combined}
\end{figure}

\gfig{fig:chi:combined} shows the combined sensitivities
at the future $e^+ e^-$ colliders as previously shown in
\gfig{fig:chi:monoa:unpolbeams} and
\gfig{fig:chi:epx:unpolbeams}. Since beam polarizations
are considered only for ILC, the sensitivity combination
is illustrated with unpolarized configurations in
\gtab{tab:mechs} for equal comparison. With relatively
smaller sensitivity enhancement from the $e^+ e^- \met$
search at the low-energy CEPC and ILC, the combined
results there are dominated by the mono-photon search
in \gfig{fig:chi:monoa:unpolbeams}. For CLIC, the
improvement of adding the $e^+ e^- \met$ search is
quite significant by increasing the limit from around
$(6 \sim 7)$\,TeV to $(30 \sim 50)$\,TeV. The mono-photon
search prevails at CEPC and ILC while $e^+ e^- \met$ dominates
at CLIC. And $e^+ e^- \met$ at ILC with beam polarizations
can help to disentangle various signal operators.

\section{Non-Collider Constraints for the Fermionic Absorption DM}
\label{sec:sec5}

As mentioned earlier, the collider search covers
the absorption operators with generic dark fermions including
the genuine fermionic absorption DM in the sub-MeV mass
range. Nevertheless,
a DM candidate needs to satisfy more constraints.
Since the effective operators in \geqn{eq:operators}
involve neutrinos and the charged electron, various
constraints can arise from the DM relic density
determined in the early Universe, the astrophysical
$X$-ray and $\gamma$ ray ($X/\gamma$-ray) observations nowadays, and the
direct detection experiments on our Earth. Although
these constraints have been systematically surveyed
in \cite{Ge:2022ius}, some updates are necessary.
With these non-collider constraints being already quite strong,
the DM search at the future lepton colliders
has its own advantage of significantly improving the
sensitivities even for very light DM as summarized in
\gfig{fig:sens:cosmo:collider}.
To make easy comparison with the existing
constraints, the collider sensitivities are also
shown in terms of $\sigma_{\chi e} v_\chi$,
\begin{eqnarray}
  \sigma_{\chi e} v_\chi
\approx 
  \frac 1 {\Lambda^4}
  { m_\chi^2 (2m_e+m_\chi)^2 (2m_e^2 + 4 m_e m_\chi + 3m_\chi^2 ) \over 32\pi (m_e+m_\chi)^4},
\end{eqnarray}
which is a function of $m_\chi$ and $\Lambda_i$
but independent of the DM velocity $v_\chi$
\cite{Ge:2022ius}.

\subsection{DM Overproduction in the Early Universe}

With highly suppressed coupling strength from cosmology,
astrophysics, and direct detection, the fermionic DM
needs to be produced through the freeze-in mechanism
at temperature around 1\,MeV \cite{Ge:2022ius}. The
production happens via the $s$-channel $e^+ e^-$ pair
annihilation $e^- e^+ \to \nu \bar \chi / \bar \nu \chi$
or the $t$-channel scatterings $e^\pm \nu \to e^\pm \chi$
and $e^\pm \bar \nu \to e^\pm \bar \chi$. Although the
inverse processes can also happen to reduce the DM
amount, its contribution is suppressed by the DM phase
space distribution function $f_\chi$ which is still
in the progress of being built. For completeness,
both contributions are included in
the Boltzmann equation that governs the DM number
density $n_\chi$ evolution, 
\begin{align}
{d n_\chi \over dt }+ 3 H n_\chi
= &
\int d\Pi_\chi d\Pi_\nu d\Pi_{e^-}d\Pi_{e^+}(2\pi)^4\delta(p_{e^+}+p_{e^-}-p_\chi - p_\nu)
\\
& \times
\left[ |{ \cal M}|_{e^-e^+\to \nu\chi}^2
f_{e^+}f_{e^-}(1-f_\nu)(1-f_\chi)
-
|{ \cal M}|_{\nu\chi\to e^-e^+ }^2f_\nu f_\chi(1-f_{e^+} )(1-f_{e^-})
\right]
\nonumber
\\
+ & 
2\int d\Pi_{e^-} d\Pi_\nu  d\Pi_{e^-}'  d\Pi_\chi  (2\pi)^4\delta(p_{e^-}+p_\nu -p_{e^-}' -p_\chi  )
\nonumber
\\
& \times
\left[ 
|{ \cal M}|_{e^- \nu \to e^- \chi}^2 f_{e^-}f_\nu (1-f_{e^-}' )(1-f_\chi) 
- |{ \cal M}|_{e^- \chi \to e^- \nu}^2 f_{e^-}' f_\chi (1-f_{e^-} )(1-f_\nu)
\right].
\nonumber
\label{eq:Bolzeq}
\end{align} 
The first integration comes from the electron-positron annihilation
while the second one from the $e^\pm \nu$ scattering. For simplicity,
we first focus on the DM particle $\chi$ while its anti-particle
has the same evolution behavior and hence the same number density,
$n_\chi = n_{\bar \chi}$. With $\chi$ appearing in the final state
of the forward scattering and the initial state of the backward
scattering, only $\nu$ needs to be taken into consideration.
Note that the phase space integration
$d\Pi_i \equiv d^3p_i / 2 E_i (2 \pi)^3$ applies for not just the
final-state particles but also their initial counterparts. Inside
the square brackets, the first term accounts for the forward reaction
while the second one for its inverse. We can see the suppression
by the DM phase space distribution $f_\chi$ appears only in the
second terms for the reversed backward reactions. Since both electron
and positrion can contribute to the $t$-channel scatterings, a
factor of 2 has been assigned to account for their identical
contributions.

\begin{figure}[t]
\centering
\includegraphics[width=0.8\textwidth]{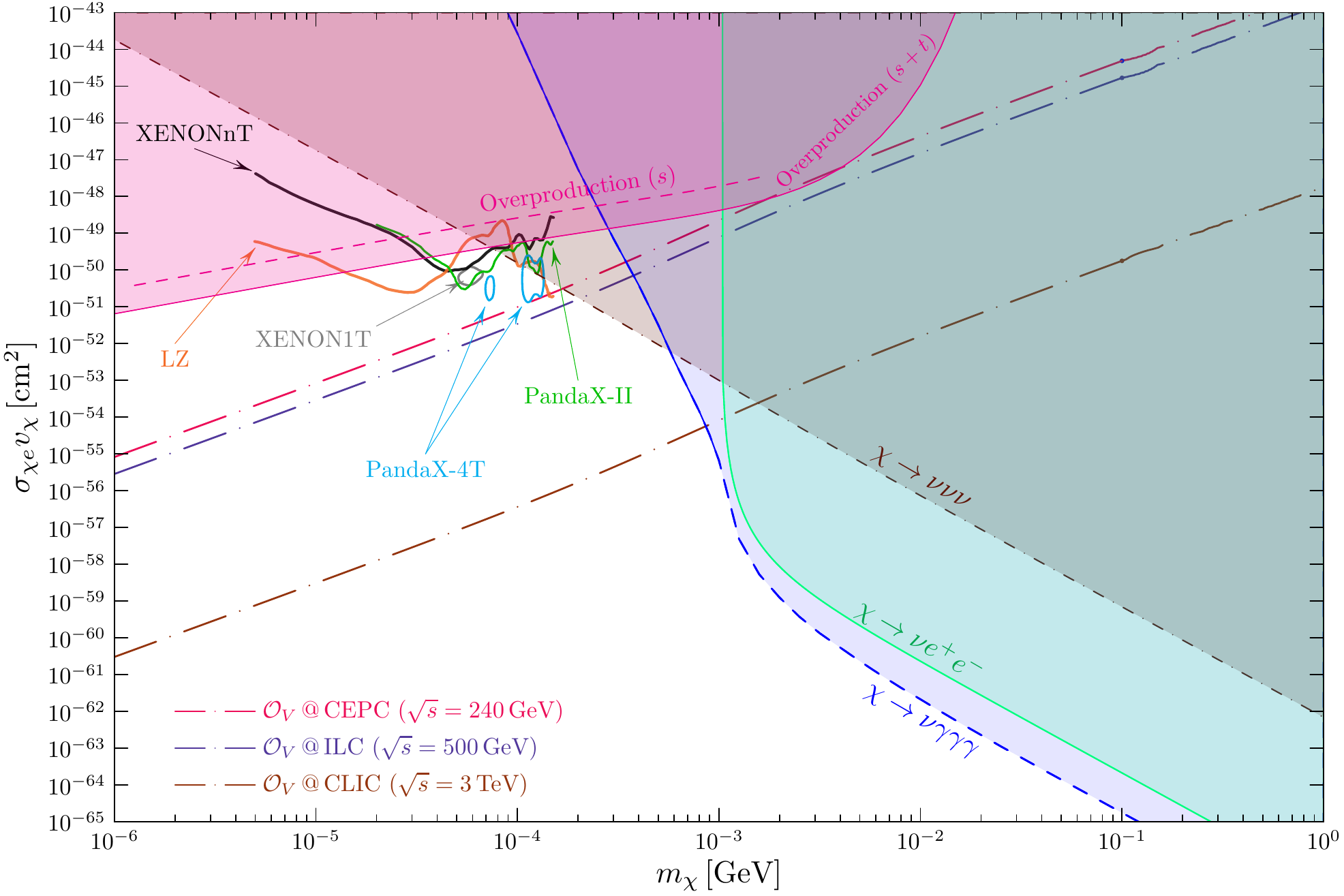}
\caption{\it The updated constraints on the fermionic
absorption DM from direct detection experiments
(PandaX-II, PandaX-4T, XENONnT, and LZ), cosmology,
and astrophysics. For comparison, the projected
sensitivities at the future lepton colliders (CEPC, ILC,
and CLIC) are shown as long dot-dashed lines.
}
\label{fig:sens:cosmo:collider}
\end{figure}

The above Boltzmann equation can be solved in terms of
the DM yield $Y \equiv n_\chi / s(T)$ as ratio of
the DM number density $n_\chi$ and the entropy density
$s(T)$.  We follow the procedures in \cite{Ge:2022ius}:
1) neglecting the Pauli blocking effect $1 - f_i \approx 1$
and the reversed reaction contributions that
are suppressed by the DM phase space distribution
$f_\chi$;
2) approximating the phase space distributions of other
particles (electron, positron, and neutrino) by their
thermal equilibrium Maxwell-Boltzmann distributions
$f_{\rm MB} = e^{- E/T}$; 3) replacing the scattering
amplitudes in \geqn{eq:Bolzeq} by the corresponding
cross sections. The final DM yield takes the following
compact form,  
\begin{align}
  Y(T)
& =
  {45M_{\rm P} \over 16\pi^6}
\int_T^{T_{\rm max}} {d \tilde T \over 1.66 \sqrt{g_*(\tilde T)} \tilde g_{*s}(\tilde T)  \tilde T^5 } 
\\
& \times 
\left[ 
\int_{s_a}^\infty ds (s-4m_e^2)\sqrt{s}K_1\left({\sqrt{s}\over \tilde  T}\right)\sigma_{e^-e^+}
+ \int_{s_b}^\infty ds 
{(s -m_e^2)^2 \over \sqrt{s}}K_1\left({\sqrt{s}\over \tilde  T}\right) \sigma_{e\nu}
\right],
\nonumber
\end{align} 
where $s_a \equiv {\rm max}(4m_e^2,m_\chi^2)$ and
$s_b \equiv (m_e + m_\chi)^2$. The maximal temperature
$T_{\rm max} = 1\,\rm MeV$ is the starting point of
the DM production. In addition, $g_*(T)$ and
$g_{*s}(T)$ are the relativistic degrees of freedom
associated with the energy and entropy densities in
the early Universe, respectively. Note that
$\tilde g_{*s} \equiv g_{*s} / \left[ 1 + (T / 3 g_{*s}) (d g_{*s} / d T) \right]$ to account for the temperature
evolution. The Planck mass $M_P$ can trace back to the
Newton constant and $K_1$ is the first modified Bessel
function of the second kind. While the pair annihilation 
cross sections $\sigma_{e^- e^+}$ can be found in \cite{Ge:2022ius},
the $t$-channel scattering ones $\sigma_{e\nu}$ are,
\begin{subequations}
\begin{align}
\sigma_{e\nu}^S 
& =
{t_-\left[ t_+^2 - 2(4m_e^2 + m_\chi^2) t_+ + 16 m_e^2 m_\chi^2
+ {t_-^2\over 3}\right] \over 64\pi (s-m_e^2)^2} {1 \over \Lambda^4}, 
\\
\sigma_{e\nu}^P 
& = 
{t_-\left( t_+^2 - 2 m_\chi^2 t_+ 
+ {t_-^2\over 3}\right) \over 64\pi (s-m_e^2)^2} {1 \over \Lambda^4} ,
\\
\sigma_{e\nu}^V 
& = 
{t_-\left\{ t_+^2 + 2(2s - m_\chi^2) t_+ 
+8 [ (s-m_e^2)^2  - m_\chi^2 s   ]
+ {t_-^2\over 3}\right\} \over 32\pi (s-m_e^2)^2} {1 \over \Lambda^4}, 
\\
\sigma_{e\nu}^A 
& = 
{t_-\left\{ t_+^2 + 2(2s - 4 m_e^2 - m_\chi^2) t_+ 
+8 [ (s-m_e^2)^2  - m_\chi^2 (s - 2m_e^2)   ]
+ {t_-^2\over 3}\right\} \over 32\pi (s-m_e^2)^2} {1 \over \Lambda^4}, 
\\
\sigma_{e\nu}^T 
& = 
{t_-\left\{ t_+^2 + 2( 4 s - 2 m_e^2 -  m_\chi^2) t_+ 
+8 [ 2(s-m_e^2)^2  - m_\chi^2 (2 s - m_e^2)   ]
+ {t_-^2\over 3}\right\} \over 8 \pi (s-m_e^2)^2} {1 \over \Lambda^4},
\end{align}
\end{subequations}
with
\begin{eqnarray}
  t_-
\equiv 
 {(s-m_e^2)\sqrt{(s-m_\chi^2 - m_e^2)^2 -4m_\chi^2 m_e^2} \over s}, 
\quad
  t_+
\equiv
- {(s-m_e^2)^2 - (s+m_e^2)m_\chi^2\over s}.
\end{eqnarray}

The generated DM yield can be converted into the DM relic
density, $\Omega_{\chi} = 2 m_\chi Y(T_0) s(T_0)/\rho_c$,
with $\rho_c$ being the critical density in the current
Universe. Note that the overall factor 2 accounts for the
identical contributions from $\chi$ and $\bar \chi$. Since
the amount of DM has already been measured quite precisely,
the DM relic density must not exceed the observed value
$\Omega_{\rm dm} h^2\approx 0.12$ \cite{Planck:2018vyg}.
This DM overproduction requirement puts stringent constraint
on the coupling strength. In \gfig{fig:sens:cosmo:collider},
the dashed magenta line reproduces the constraint with only
the $s$-channel $e^+e^-$ annihilation while the new constraint
with also the $t$-channel $e^\pm \nu$ scattering is given in
the magenta filled region. Including the $t$-channel contribution
improves the constraint by almost an order of magnitude.
The $t$-channel contribution cannot be ignored.

\subsection{DM Decay and Astrophysical $X/\gamma$ Ray Observations}

In the presence of coupling with charged electron and
neutrino, the DM can decay into light SM particles including
not just the massless photon and neutrinos but also electron
and positron if the DM is heavy enough, $m_\chi > 2 m_e$.
We provide updated calculation and constraints for this
heavy DM region. Due to space limitation, only results for
the vector case are shown in \gfig{fig:sens:cosmo:collider}
and our discussions below focus on this particular case.

{\it $\pmb{\chi \to \nu\nu\nu}$}:
For DM mass $m_\chi \lesssim 2m_e$, the dominant
decay mode is the invisible channel $\chi \to \nu\nu\nu$.
Since neutrinos are relativistic degrees of freedom,
the DM decay in the early Universe will inject radiative
energy to influence the cosmic evolution. The constraint
is obtained in \cite{Ge:2022ius} and extended above
$2 m_e$ to 1\,GeV as the dashed brown line in
\gfig{fig:sens:cosmo:collider}. Although the branching
ratio of this channel becomes suppressed in the presence
of a new channel $\chi \rightarrow e^+ e^- \nu$ for
$m_\chi > 2 m_e$, the cosmological expansion is sensitive
to the decay width instead and hence the extension of
the constraint curve is smoothly extended.

{\it $\pmb{\chi \to \nu\gamma\gamma\gamma}$}:
For the vector operator with DM mass $m_\chi \lesssim 2m_e$,
the dominant visible decay mode is the four-body channel
$\chi \to \nu \gamma\gamma\gamma$. The constraint is shown
as blue solid and dashed lines with division at $m_\chi = 2m_e$
using the same observational data summarized in \cite{Ge:2022ius}.
It is obtained by requiring that the predicted
photon flux from galactic and/or extragalactic DM decay 
does not exceed the diffuse $X/\gamma$-ray background collected 
by the astrophysical observations (Insight- HXMT, NuSTAR, HEAO-1, and INTEGRAL).
We can see the scaling behaviors with $m_\chi$ below and
above $2 m_e$ are quite different. For the light DM case,
$\Gamma_{\chi \to \nu \gamma\gamma\gamma} \propto
m_\chi^{13}/m_e^8\Lambda^4$ \cite{Ge:2022ius} while
for the heavy one $\Gamma_{\chi \to \nu \gamma\gamma\gamma}
\propto m_\chi^5/\Lambda^4$. On the other hand, the vertical
axis variable scales as
$\sigma_{\chi e} v_\chi \propto m_\chi^2/\Lambda^4$.
Putting things together, $\sigma_{\chi e}v_\chi \propto m_\chi^{-11}\,(m_\chi^{-3})$ for $m_\chi \ll 2m_e\,(m_\chi \gg 2m_e$),
respectively. That is why the blue curve has much larger
slope for light DM.

{\it $\pmb{\chi \to \nu e^+ e^-}$}:
For the DM mass beyond the electron-positron pair production
threshold, $m_\chi > 2 m_e$, the dominant visible decay mode
is $\chi \to \nu e^+ e^-$. The decay
width of $\chi \to \nu e^+ e^-$ due to the vector operator
${\cal O}_V$ takes the form as,
\begin{eqnarray}
  \Gamma_{\chi\to\nu e e}^V
=
 {m_\chi^5 \over 6144\pi^3 \Lambda^4}
 \left[
 (16 - 40 \eta + 18 \eta^2 -9 \eta^3)\sqrt{1- \eta}
 + 3(8 - 3 \eta) \eta^3 \tanh^{-1}(\sqrt{1-\eta})
 \right],
\end{eqnarray}
where $\eta \equiv 4 m_e^2/ m_\chi^2$. Requiring stable DM on
the cosmological time scale, with a DM lifetime $\tau_\chi$
larger than the age of the Universe,
$\tau_\chi \approx 1/\Gamma_{\chi\to\nu e e}
> 13.8\,{\rm Gyr}$, puts a conservative constraint as shown
with the solid green line in \gfig{fig:sens:cosmo:collider}.
Even though the $\chi \to \nu e^+ e^-$ channel dominates over
$\chi \to \nu \gamma\gamma\gamma$, this constraint is still
weaker by an order of magnitude or so. 

For completeness, we also list the
$\chi \to \nu e e $ decay width with $m_\chi \geq 2m_e$
from other operators, 
\begin{align}
  \Gamma_{\chi\to\nu e e}^S
& =
  {m_\chi^5 \over 24576 \pi^3 \Lambda^4} 
\left[
 (16 - 88 \eta  - 42 \eta^2 + 9 \eta^3)\sqrt{1- \eta}
 + 3(48 - 16 \eta + 3 \eta^2) \eta^2 
 \tanh^{-1}(\sqrt{1-\eta})
\right], 
\nonumber
\\
  \Gamma_{\chi\to\nu e e}^P
& =
  {m_\chi^5 \over 3072\pi^3 \Lambda^4} 
\left[
- 5( 2 - \eta)
+(16 - 16 \eta + 5 \eta^2) \eta^2 {}_2F_1\left({3\over 2},3; {5 \over 2};1-\eta\right)
\right](1- \eta)^{3\over 2},
 \nonumber
\\
\Gamma_{\chi\to\nu e e}^A
& =
  {m_\chi^5 \over 6144\pi^3 \Lambda^4}
 \left[
 (16 - 72 \eta -22 \eta^2 +3 \eta^3)\sqrt{1- \eta}
 + 3(32 - 8 \eta + \eta^2) \eta^2 
 \tanh^{-1}(\sqrt{1-\eta})
 \right], 
 \nonumber
 \\
\Gamma_{\chi\to\nu e e}^T
& =
 {m_\chi^5 \over 1024\pi^3 \Lambda^4} 
  \left[
 (16 - 56 \eta -2 \eta^2 - 3 \eta^3)\sqrt{1- \eta}
 +  3(16 -  \eta^2) \eta^2 \tanh^{-1}(\sqrt{1-\eta})
 \right], 
\end{align}
where ${}_2F_1(a,b;c;z)$ is the hypergeometric function.
A factor of 2 is included to account for both neutrino
and anti-neutrino final states. Other visible and
invisible decay modes have been explored in
our previous work \cite{Ge:2022ius}. For scalar,
pseudo-scalar, and axial-vector operators, similar to 
\gfig{fig:sens:cosmo:collider}, the collider searches
are also sensitive to light DM while the DM
overproduction and decay constraints are stronger for
the heavy counterpart. However, for the tensor case, the collider
search is always worse than the DM visible decay mode
$\chi\to \nu\gamma$ due to its large decay width. 

For dark fermion with $m_\chi \gg 2m_e$, we can estimate
the dark fermion lifetime from the vector-type operator,  
\begin{align}
\tau_\chi^V  \approx \Gamma_{\chi \nu ee}^{V,-1}\approx 
\left( {m_\chi^5 \over 384 \pi^3 \Lambda^4} \right)^{-1}
= 7.83\times 10^{-19}\,{\rm s} 
\left(\Lambda \over 1\,{\rm TeV} \right)^4 
\left(100\,{\rm GeV} \over m_\chi \right)^5,
\end{align}
from which the others are simple scalings,
$\tau_\chi^{(S,P,A,T)}\approx (4,4,1,1/6)\tau_\chi^V$.
With strong boost effect for an energetic particle
produced at collider, the decay length scales as,
\begin{equation}
 L_{\chi}^{V} \approx 2.349 \times 10^{-10} {\rm m} \times
 \frac{1}{2} \frac{\sqrt{s}}{m_\chi} \bigg(1 + \frac{m_\chi^2}{s} \bigg)  
 \left(\Lambda \over 1\,{\rm TeV} \right)^4 
\left(100\,{\rm GeV} \over m_\chi \right)^5.
\end{equation}
It can reach $L_X^{V} \sim 10^{-4}\,\rm m$ at CEPC
for $m_\chi = 10\,$GeV and $\Lambda = 1$\,TeV.
In other words, a heavy dark fermion can decay inside the
detector and with a displaced vertex for $\mathcal O$(GeV)
masses. But sub-GeV dark fermions contribute as missing
energy. So more information can be extracted from its
decay products to obtain even stronger constraint on
the relevant absorption operators for at least
GeV dark fermions.
To be conservative, we only consider the missing energy
search and leave more detailed studies for further study.

\subsection{Direct Detection}

The direct detection experiments are sensitive to
the sub-MeV scale DM that leaves $\mathcal{O}(10)$\,keV
electron recoil. We use XENON1T \cite{XENON:2020rca},
PandaX-II \cite{PandaX-II:2020udv},
PandaX-4T \cite{PandaX:2022ood},
LZ \cite{LZ:2022ufs},
and XENONnT \cite{XENON:2022ltv} to give an updated constraint.
In addition to the fermionic DM absorption signal,
the background estimations are also taken into
consideration with a nuisance parameter \cite{Ge:2021snv}. 
Using the analytical $\chi^2$ analysis
\cite{Ge:2012wj,Ge:2016zro,Ge:2022ius},
we obtain the $95\%$ C.L. sensitivity
curves/contours for XENON1T (gray), PandaX-II (green),
PandaX-4T (blue), LZ (orange), and XENONnT (black)
as shown in \gfig{fig:sens:cosmo:collider}. 
While XENON1T and PandaX-4T give contours, the
results from PandaX-II, LZ, and XENONnT are exclusion
limits at 95\% C.L.

\section{Discussions and Conclusions}
\label{sec:conclusion}

The collider searches can probe not just the genuine
DM but also other dark sector particles.
We focus on the absorption operators that couple
a dark fermion with neutrino and the charged
electron/positron. Although such operators have been
highly constrained by
the cosmological and astrophysical observations as
systematically explored in our previous
phenomenological paper \cite{Ge:2022ius},
the future $e^+ e^-$ colliders can provide even better
sensitivities particularly in the sub-MeV
mass range. The future lepton colliders
(CEPC, ILC, and CLIC) can even surpass the current DM
direct detection experiments (PandaX and XENON1T).
In other words, the fermionic absorption
DM is of great interest to not just the direct detection
but also the collider search. While the indirect
detection prevails for heavy DM and the direct detection
for the keV scale, the collider search dominates at
the light mass range and is hence of great advantage.
So the three independent searches are still complementary
to each other for the absorption operator scenario.
Comparable feature happens for the cosmological
limits from the DM overproduction requirement but the
sensitivity is much weaker than the collider one.

\section*{Acknowledgements}

SFG and KM would like to thank Ning Zhou, Hai-Jun Yang,
and Manqi Ruan for useful discussions.
The authors are supported by the National Natural Science
Foundation of China (12090060, 12090064, 12375101, and 12305110) and
the SJTU Double First Class start-up fund (WF220442604).
SFG is also an affiliate member of Kavli IPMU, University of Tokyo.
KM is supported by the Innovation Capability Support Program of Shaanxi (Program No. 2021KJXX-47).
XDM is also supported by Guangdong Major Project of Basic and Applied Basic Research (No. 2020B0301030008).

\end{document}